\documentclass{aa}  %print version
\usepackage{graphicx}
\usepackage{txfonts}
\usepackage{natbib}
\usepackage{color}
\usepackage{tabularx}
\usepackage{tabu}
\pdfoutput=1 %for arXiv submission
\usepackage{amsmath,amstext}
\usepackage{nicefrac}
\usepackage{multirow}
\usepackage{tablefootnote}
\usepackage[colorlinks=true,allcolors=blue]{hyperref}
\usepackage{float}

\begin{document} 

\title{Microlensing of the broad emission lines in 27 gravitationally lensed quasars}
\subtitle{Broad line region structure and kinematics}

\author{C. Fian\inst{1}, E. Mediavilla\inst{2,3}, V. Motta\inst{4}, J. Jim\'enez-Vicente\inst{5,6}, J. A. Mu\~noz\inst{7,8}, D. Chelouche\inst{9}, A. Hanslmeier\inst{10}}

\institute{School of Physics and Astronomy and Wise Observatory, Raymond and Beverly Sackler Faculty of Exact Sciences, Tel-Aviv University, Tel-Aviv, Israel\label{inst1} \and Instituto de Astrof\'{\i}sica de Canarias, V\'{\i}a L\'actea S/N, La Laguna 38200, Tenerife, Spain\label{inst3} \and Departamento de Astrof\'{\i}sica, Universidad de la Laguna, La Laguna 38200, Tenerife, Spain\label{inst4} \and Instituto de F\'{\i}sica y Astronom\'{\i}a, Universidad de Valpara\'{\i}so, Avda. Gran Breta\~na 1111, Playa Ancha, Valpara\'{\i}so 2360102, Chile \and Departamento de F\'{\i}sica Te\'orica y del Cosmos, Universidad de Granada, Campus de Fuentenueva, 18071 Granada, Spain \and Instituto Carlos I de F\'{\i}sica Te\'orica y Computacional, Universidad de Granada, 18071 Granada, Spain \and Departamento de Astronom\'{i}a y Astrof\'{i}sica, Universidad de Valencia, E-46100 Burjassot, Valencia, Spain \and Observatorio Astron\'{o}mico, Universidad de Valencia, E-46980 Paterna, Valencia, Spain \and Haifa Research Center for Theoretical Physics and Astrophysics, University of Haifa, Haifa, Israel\label{inst2} \and Institute of Physics (IGAM), University of Graz, Universit{\"a}tsplatz 5, 8010, Graz, Austria}
%\date{Received xxxxxxx; accepted xxxxxxx}

%%%%%%%%%%%%%%%%%%%%%%%%%%%%%%%%%%%%%%%%%%%%%%%%%%%%%%%%%%%%%%%%%%%%%%%%%%%%%%%%%%%%%%%%%%%%%%
%%%%%%%%%%%%%%%%%%%%%%%%%%%%%%%%%%%%%%%%%%%%%%%%%%%%%%%%%%%%%%%%%%%%%%%%%%%%%%%%%%%%%%%%%%%%%%
\abstract
{}
{We aim to study the structure and kinematics of the broad line region (BLR) of a sample of 27 gravitationally lensed quasars with up to five different epochs of observation. This sample is composed of $\sim$100 spectra from the literature plus 22 unpublished spectra of 11 systems.}
{We measure the magnitude differences in the broad emission line (BEL) wings and  statistically model the distribution of microlensing magnifications to determine a maximum likelihood estimate for the sizes of the C IV, C III], and Mg II emitting regions.}
{The BELs in lensed quasars are expected to be magnified differently owing to the different sizes of the regions from which they originate. Focusing on the most common BELs in our spectra (C IV, C III], and Mg II), we find that the low-ionization line Mg II is only weakly affected by microlensing. In contrast, the high-ionization line C IV shows strong microlensing in some cases, indicating that its emission region is more compact. Thus, the BEL profiles are deformed differently depending on the geometry and kinematics of the corresponding emitting region. We detect microlensing in either the blue or the red wing (or in both wings with different amplitudes) of C IV in more than 50$\%$ of the systems and find outstanding asymmetries in the wings of QSO 0957+561, SDSS J1004+4112, SDSS J1206+4332, and SDSS J1339+1310. This observation indicates that the BLR is, in general, not spherically symmetric and supports the existence of two regions in the BLR, one insensitive to microlensing and another that only shows up when it is magnified by microlensing. Disregarding the existence of these two 
regions, our estimate for Mg II, $R_{1/2} = 67.3_{-15.7}^{+3.8} \sqrt{M/M_\odot}$ light-days, is in good agreement with previous results from smaller samples, while we obtain smaller sizes for the C III] and CIV lines, $R_{1/2} = 31.0_{-4.0}^{+1.9} \sqrt{M/M_\odot}$ light-days and $R_{1/2} = 15.5_{-3.9}^{+0.8} \sqrt{M/M_\odot}$ light-days, respectively.}
{}
\keywords{quasars: emission lines --- gravitational lensing: micro --- quasars: general --- quasars: supermassive black holes}

\titlerunning{Broad Line Region Structure and Kinematics}
\authorrunning{Fian et al.} 
\maketitle

\section{Introduction}
Gravitational lensing is a powerful tool for studying the structure of lensed quasars (\citealt{Pooley2007,Mosquera2009,Blackburne2011,Mosquera2011,Braibant2014}). Since the discovery of multiply lensed quasars, anomalies in the flux ratios between the images have been found, which were thought to be associated with different phenomena: a complex mass distribution of the lens galaxy, dust extinction, dark matter substructure, and microlensing. Quasar microlensing, produced by compact objects (i.e., stars) in the lens galaxy, is one of the most likely explanations (\citealt{Chang1979,congdon2005,wambsganss2006}). They differentially magnify components of the quasar emission regions, leading to time- and wavelength-dependent changes in the flux ratios of the images (\citealt{wambsganss2006,Abajas2002}). Microlensing is size sensitive, with smaller source regions showing larger magnifications. Regions with sizes comparable to the angular Einstein radius, $R_E$, or smaller are expected to be significantly magnified (\citealt{Wambsganss1998,wambsganss2006,Schmidt2010}). \\

Strong BELs are characteristic of many active galactic nuclei and are produced over a wide range of distances from the central source (see, e.g., \citealt{Sulentic2000}). The narrow line region (NLR) is typically large enough to be insensitive to microlensing by solar mass objects. Also, the BELs in gravitationally lensed quasars are expected to be less affected by microlensing than the continuum (see 
early studies by \citealt{Nemiroff1987,Schneider1990}). To be precise, microlensing affects the broad wings of the emission lines that correspond to regions of high velocities, whereas the line cores are supposed to be insensitive to microlensing as they arise from regions spatially much more extended than the region contributing to the wings (\citealt{Abajas2002,Richards2004,Lewis2004,Gomez2006,Guerras2013}). For this reason, we use the line cores as reference to set the baseline for no microlensing. The wings of low-ionization lines are less affected by microlensing, indicating that they arise from a considerably larger region than the high-ionization lines (\citealt{Guerras2013}). The study of changes induced in the line profiles by microlensing impose significant constraints on the geometry and kinematics of the BLR and on its dependence on the ionization degree.\\

In our previous work (\citealt{Fian2018}) we compared the emission line profiles of pairs of images from a sample of 11 lensed quasars for which archival spectroscopy was available in more than one epoch. We were able to measure microlensing of $\sim$0.11 mag on average for the low-ionization lines, which allowed us to produce estimates for the BLR size of $r_s = 50.3_{-14.0}^{+30.4}$ light-days. Here, we improve these results by extending the study to a sample of 27 lensed quasars with up to five different epochs of observation and by estimating the size of the emission region for each emission line. Our data consist of microlensing measurements based on the comparison between the flux ratios of the wings between pairs of images of around 100 high-ionization lines (C IV) and nearly 200 low-ionization lines (C III] and Mg II). Prior to that, we match the cores of the emission lines to provide an unmicrolensed baseline that also removes the effects of dust extinction. In this way, constraints on the size of the BLR and on its kinematics can be obtained. To our knowledge, no similar long time baseline spectroscopic study of gravitational lenses has ever been carried out before, so these data are a new and unique probe of the BLR structure.\\

The paper is organized as follows. In Section \ref{sec2} we describe the spectra collected from the literature as well as the observation and data reduction of the new, unpublished spectra. Section \ref{sec2} also includes the analysis of the microlensing signal observed in the BELs. In Section \ref{sec3} the main results are shown. In Section \ref{sec4} we use these microlensing estimates to derive constraints on the typical size of the BEL emitting regions of an average quasar. Section \ref{sec5} is devoted to the kinematics of the BLR. In Section \ref{sec6} we estimate the average mass of the central supermassive black hole (SMBH) of our lensed quasar sample using the previously obtained sizes and kinematic information. We discuss and summarize our results in Section \ref{sec7}.

%%%%%%%%%%%%%%%%%%%%%%%%%%%%%%%%%%%%%%%%%%%%%%%%%%%%%%%%%%%%%%%%%%%%%%%%%%%%%%%%%%%%%%%%%%%%
%%%%%%%%%%%%%%%%%%%%%%%%%%%%%%%%%%%%%%%%%%%%%%%%%%%%%%%%%%%%%%%%%%%%%%%%%%%%%%%%%%%%%%%%%%%%
\section{Data analysis}\label{sec2}
\subsection{Data and observations}
In \citet{Fian2018} we collected from the literature rest frame UV spectra of lensed quasars. In this work we add unpublished spectra taken with the 4.2m William Herschel Telescope (WHT), located at the Roque de los Muchachos Observatory in La Palma (Canary Islands, Spain). We obtained ISIS\footnote{Intermediate-dispersion Spectrograph and Imaging System} long-slit spectroscopy covering the full optical range for 11 systems for which we had spectra available in previous epochs. The targets were observed with both arms of the ISIS instrument using the gratings R300B and R316B for the blue and red arm, respectively. With this setup, we were able to observe the full range between 3500 and 9000 \AA\ with a spectral sampling of 0.9 \AA/pix. This spectral range includes several typical high- and low-ionization lines of quasars (C IV, C III], and Mg II). The data were reduced using standard IRAF routines for long-slit 2D spectroscopic data. These included bias subtraction, flat field and illumination correction, cosmic ray removal, wavelength calibration, background subtraction, flux calibration, and extraction of the 1D spectra. The data from the literature were already fully reduced. To avoid cross-contamination between the spectra of closely spaced image pairs, we fit two Gaussians to the reduced data in order to separate them. To provide a bigger statistical sample and emphasize trends, we also included nine lens systems with single-epoch spectra. In total, we gathered a sample of 32 pairs of quasar images in 27 lens systems with up to five epochs of observation. The superpositions of the emission lines corresponding to different images and epochs for each system are shown in Figures \ref{em1}-\ref{em4}. Information about the lens systems and references are summarized in Table \ref{table1}. The following data are provided: name of the object, number of images, number of epochs, observing date, facility, and reference. 

\begin{table*}
\setlength{\tabcolsep}{4.2pt}
\renewcommand{\arraystretch}{0.95}
\caption{Database of lensed quasar spectra.}
\label{table1}
\centering
\begin{tabular}{lcccccccc}
\hline\hline
Object & Image & Epoch & Date & C IV & C III] & Mg II & Facilities & References\\
\hline
\multirow{3}{*}{HE 0047-1756} & \multirow{3}{*}{A,B} & I & 2002 Sep 04 & x & x & - & Magellan & \citealt{Wisotzki2004}\\
	& & II & 2005 Jul 18 & - & x & x & VLT & \citealt{Sluse2012}\\
    & & III & 2008 Jan 13 & - & x & x & Magellan & \citealt{Rojas2014}\\ \hline \vspace*{-3mm}\\	
	\multirow{2}{*}{Q 0142-100} & \multirow{2}{*}{A,B} & I & 2006 Aug 15 & x & x & - & VLT & \citealt{Sluse2012}\\
	& & II & 2008 Jan 12 & x & - & - & MMT & Motta (private communication)\\ \hline \vspace*{-3mm}\\ 	
    SDSS J0246-0825 & A,B & I & 2006 Aug 22 & - & x & x & VLT & \citealt{Sluse2012}\\ \hline \vspace*{-3mm}\\	
	\multirow{4}{*}{HE 0435-1223} & A,B & I & 2002 Sep 05 & x & x & - & CAO & \citealt{Wisotzki2003}\\ \cline{2-9} \vspace{-3mm}\\
	& B,D & II & 2004 Oct-Nov & - & x & x & VLT & \citealt{Eigenbrod2007}\\ \cline{2-9} \vspace{-3mm}\\
	& A,C,D & III & 2007 Dec 10 & x & x & x & Magellan & \citealt{Motta2017} \\ \cline{2-9} \vspace*{-3mm}\\	
    & A,B,C,D & IV & 2008 Jan 12 & x & x & x & MMT & \citealt{Motta2012}\\ \hline \vspace{-3mm}\\
     HE 0512-3329 & A,B & I & 2001 Aug 13 & x & x & - & HST & \citealt{Wucknitz2003}\\ \hline \vspace*{-3mm}\\	
 	\multirow{2}{*}{SDSS J0806+2006} & \multirow{2}{*}{A,B} & I & 2005 Apr 12 & - & x & x & APO & \citealt{Inada2006}\\
 	& & II & 2006 Apr 22 & - & x & x & VLT & \citealt{Sluse2012}\\ \hline \vspace*{-3mm}\\	
 	HS 0818-1227 & A,B & I & 2008 Jan 12 & x & - & - & MMT & \citealt{Motta2012} \\ \hline \vspace*{-3mm}\\
 	\multirow{2}{*}{SBS 0909+532} & \multirow{2}{*}{A,B} & I & 2003 Mar 07 & x & x & x & HST & \citealt{Mediavilla2005}\\ 
     & & II & 2016 Mar 17 & x & x & x & WHT & Mediavilla, Jimenez-Vicente, Fian \\ \hline \vspace*{-3mm}\\	
 	\multirow{2}{*}{SDSS J0924+0219} & \multirow{2}{*}{A,B} & I & 2005 Jan 14 & - & x & x & VLT & \citealt{Eigenbrod2006}\\ 
     & & II & 2016 Mar 17 & x & x & x & WHT & Mediavilla, Jimenez-Vicente, Fian \\ \hline \vspace*{-3mm}\\	
     \multirow{3}{*}{FBQ 0951+2635} & \multirow{3}{*}{A,B} & I & 1997 Feb 14 & - & - & x & Keck & \citealt{Schechter1998}\\
 	& & II & 2006 Mar 31 & - & - & x & VLT & \citealt{Sluse2012}\\ 
     & & III & 2016 Mar 17 & x & x & x & WHT & Mediavilla, Jimenez-Vicente, Fian \\ \hline \vspace*{-3mm}\\	
 	\multirow{4}{*}{Q 0957+561} & A & I & 1999 Apr 15 & x & x & x & HST & \citealt{Goicoechea2005}\\ \cline{2-9} \vspace{-3mm}\\
 	& B & I & 2000 Jun 2 & x & x & x & HST & \citealt{Goicoechea2005}\\ \cline{2-9} \vspace{-3mm}\\
 	& \multirow{2}{*}{A,B} & II & 2008 Jan 12 & x & x & x & MMT & \citealt{Motta2012} \\ 
     & & III & 2016 Mar 12 & x & x & x & WHT & Mediavilla, Jimenez-Vicente, Fian \\ \hline \vspace*{-3mm}\\	
     SDSS J1001+5027 & A,B & I & 2003 Nov 20 & x & x & x & APO & \citealt{Oguri2005}\\ \hline \vspace*{-3mm}\\
 	\multirow{4}{*}{SDSS 1004+4112} & A,B,C,D & I & 2003 May 31 & x & x & x & APO & \citealt{Richards2004}\\ \cline{2-9} \vspace{-3mm}\\
 	& \multirow{3}{*}{A,B} & II & 2004 Jan 19 & x & x & - & WHT & \citealt{Gomez2006}\\ 
 	& & III & 2008 Jan 12 & x & x & x & MMT & \citealt{Motta2012}\\ 
     & & IV & 2016 Mar 11 & x & x & x & WHT & Mediavilla, Jimenez-Vicente, Fian \\ \hline \vspace*{-3mm}\\	
     Q 1017-207& A,B & I & 1996 Oct 28 & x & x & - & HST & \citealt{Surdej1997}\\ \hline \vspace*{-3mm}\\
 	\multirow{2}{*}{SDSS J1029+2623} & \multirow{2}{*}{A,B} & I & 2008 Jan 12 & x & x & - & MMT & \citealt{Motta2012} \\ 
     & & II & 2016 Mar 12 & x & x & - & WHT & Mediavilla, Jimenez-Vicente, Fian \\ \hline \vspace*{-3mm}\\	
 	\multirow{5}{*}{HE 1104-1805} & \multirow{5}{*}{A,B} & I & 1993 May 11 & x & x & - & NTT & \citealt{Wisotzki1993}\\
 	& & II & 1994 Nov 29 & x & - & - & ESO 3.6m & \citealt{Wisotzki1995} \\
 	& & III & 2008 Jan 11 & x & - & - & MMT & \citealt{Motta2012} \\
 	& & IV & 2008 Apr 07 & x & x & x & VLT & \citealt{Motta2012} \\ 
     & & V & 2016 Mar 12 & x & x & - & WHT & Mediavilla, Jimenez-Vicente, Fian \\ \hline \vspace*{-3mm}\\	
     SDSS J1138+0314 & B,C & I & 2005 May 10 & x & x & - & VLT & \citealt{Sluse2012}\\ \hline \vspace*{-3mm}\\
 	SDSS J1155+6346 & A,B & I & 2010 Sep 20 & x & x & - & HST & \citealt{Rojas2014} \\ \hline \vspace*{-3mm}\\
 	\multirow{2}{*}{SDSS J1206+4332} & \multirow{2}{*}{A,B} & I & 2004 Jun 21 & x & x & x & APO & \citealt{Oguri2005}\\ 
     & & II & 2016 Mar 12 & x & - & x & WHT & Mediavilla, Jimenez-Vicente, Fian\\ \hline \vspace*{-3mm}\\
 	SDSS J1335+0118 & A,B & I & 2005 Feb 17 & - & x & x & VLT & \citealt{Sluse2012}\\ \hline \vspace*{-3mm}\\
 	\multirow{3}{*}{SDSS J1339+1310} & \multirow{3}{*}{A,B} & I & 2013 Apr 13 & x & x & x & GTC & \citealt{Shalyapin2014}\\
 	& & II & 2014 Mar 27 & x & x & x & GTC & \citealt{Goicoechea2016}\\
 	& & III & 2014 May 20 & x & x & - & GTC & \citealt{Goicoechea2016}\\ \hline \vspace*{-3mm}\\	
     \multirow{2}{*}{SDSS J1353+1138} & \multirow{2}{*}{A,B} & I & 2005 Apr 12 & - & x & x & Keck & \citealt{Inada2006}\\ 
     & & II & 2016 Mar 17 & x & - & - & WHT & Mediavilla, Jimenez-Vicente, Fian \\ \hline \vspace*{-3mm}\\
     Q 1355-2257 & A,B & I & 2005 Mar 13 & - & x & x & VLT & \citealt{Sluse2012}\\ \hline \vspace*{-3mm}\\
     B 1422+231 & A,B & I & 2016 Mar 17 & x & - & - & WHT & Mediavilla, Jimenez-Vicente, Fian \\ \hline \vspace*{-3mm}\\
 	\multirow{2}{*}{SBS 1520+530} & \multirow{2}{*}{A,B} & I & 1996 Jun 12 & x & x & - & SAO & \citealt{Chavushyan1997}\\ 
     & & II & 2016 Mar 12 & x & - & - & WHT & Mediavilla, Jimenez-Vicente, Fian \\ \hline \vspace*{-3mm}\\
	\multirow{3}{*}{WFI 2033-4723} & A1,A2,B,C & I & 2003 Sep 15 & x & x & x & Magellan & \citealt{Morgan2004}\\ \cline{2-9} \vspace{-3mm}\\
 	& \multirow{2}{*}{B,C} & II & 2005 May 13 & - & x & x & VLT & \citealt{Sluse2012}\\ 
 	& & III & 2008 Apr 14 & - & x & x & VLT & \citealt{Motta2017} \\ \hline \vspace*{-3mm}\\	
 	\multirow{3}{*}{HE 2149-2745} & \multirow{3}{*}{A,B} & I & 2000 Nov 19 & x & x & x & VLT & \citealt{Burud2002}\\
 	& & II & 2006 Aug 04 & x & x & - & VLT & \citealt{Sluse2012}\\
 	& & III & 2008 May 07 & x & x & x & VLT & \citealt{Motta2017}\\	\hline \vspace*{-3mm}\\
\end{tabular}
\end{table*}

\subsection{Data analysis methods}\label{sec2.2}
We focus on the low-ionization lines C III] $\lambda$1909 and Mg II $\lambda$2798 and the high-ionization line C IV $\lambda$1549. For each emission line we used DIPSO in STARLINK to fit a straight line $y=a\lambda+b$ to the continuum on either side of the emission line and subtract it from the spectrum. In order to quantify the effects of microlensing on the BLR, we want to untangle microlensing from the macro-magnification produced by the lens galaxy and extinction (the last two parameters are independent of the source size). We attempt this by normalizing the continuum-subtracted spectra for all images and all epochs to match the core of the emission line defined by the flux within a narrow interval ($\pm$6\AA) centered on the peak of the line. The continuum-subtracted and core matched spectra in the wavelength regions around the C IV, C III], and Mg II emission lines can be seen in Figures \ref{em1} to \ref{em4}. In principle, the cores of the emission lines can be used as a reference that is little affected by microlensing and intrinsic variability (see \citealt{Guerras2013,Fian2018}) as they arise from a significantly larger region than the wings. Under this assumption, the ratio of the line wing fluxes between pairs of images at the same epoch or between different epochs for the same image, $F_{1wings}/F_{2wings}$, yields a measurement of the size of the emitting region. To prevent an underestimation of the microlensing in the wings we separate the line core from the wings by a buffer of $\pm$9\AA. We then can estimate the average wing emission in different wavelength intervals ($\sim$25\AA\ for C IV, $\sim$35\AA\ for CIII] and Mg II) on either side of the emission line peak, corresponding to velocity intervals of $\sim$4500 km/s for C IV, $\sim$5300 km/s for CIII], and $\sim$3600 km/s for Mg II. In those cases in which the emission line is affected by absorption lines, an integration window avoiding absorption features was chosen (see Figures \ref{em1} to \ref{em4}). We use the following statistics to calculate the magnitude difference in the wings at each wavelength $x$ between two different images/epochs ($\alpha$,$\beta$):
\begin{equation}
\Delta m_x =w_x*(\beta_x-\alpha_x), 
\end{equation}

with weights $w_x=\sqrt{<\beta_x+\alpha_x>/(\beta_x+\alpha_x)}$, selected to equalize the typical deviations of the differences. From the mean value in a given wavelength interval, $\langle \Delta m_x \rangle$, we compute the magnitude difference between images/epochs, $\Delta m = \langle \Delta m_x \rangle$, and its standard deviation $\sigma$. We did not use values with uncertainties above a given threshold ($\sigma=0.5$mag). The estimated magnitude differences between images with S/N greater than 1.5 are given for each epoch in Table \ref{table2} and similarly, the magnitude differences between epochs for each image are presented in Table \ref{table3}. There is no way to  entirely separate microlensing from intrinsic variability without having observations separated by exactly the time delay between images. However, to qualify the magnitude differences between images and/or epochs as candidates for microlensing or intrinsic variability, we used the same criteria as in \citealt{Fian2018}: (i) the S/N should be greater than 2, (ii) any difference between images is considered a candidate for microlensing, (iii) we consider as a candidate for intrinsic variability a difference between two epochs when it is present in at least two images, (iv) when neither (ii) nor (iii) apply we consider that we have insufficient information to qualify the difference, although intrinsic variability may be more likely (partial evidence of intrinsic variability). We note that in this paper the classification is solely based on the differences in the C IV, C III], and Mg II emission lines while in \citet{Fian2018} we also took into account differences in other emission features (red shelf of C IV, several Fe II and Fe II blends); therefore, some classifications may differ. The resulting classifications for this work are shown in Tables \ref{table2} and \ref{table3}.

\begin{table*}
\setlength{\tabcolsep}{9.6pt}
\renewcommand{\arraystretch}{1}
\caption{Differences between images in the C IV, C III], and Mg II lines.}
\label{table2}
\centering
\begin{tabular}{lccccccc}
\hline\hline
Emission Line & Wing & Object & Image Pair & Epoch & $\Delta$m$^{*}$ $\pm$ $\sigma^{**}$ & $\Delta$m/$\sigma$ & Classification\\
\hline \vspace{-3mm} \\ 
    \multirow{28}{*}{C IV} & \multirow{11}{*}{blue wing} & FBQ 0951+2635 & B-A & III & 0.52 $\pm$ 0.25 & 2.1 & Microlensing \\ \cline{3-8} \vspace*{-3mm}\\
    & & \multirow{6}{*}{SDSS 1004+4112} & \multirow{3}{*}{B-A} & I & 0.56 $\pm$ 0.16 & 3.4 & Microlensing \\
	& & & & III & 0.39 $\pm$ 0.07 & 5.8 & Microlensing \\ 
    & & & & IV & 0.38 $\pm$ 0.14 & 2.8 & Microlensing \\ \cline{4-8} \vspace*{-3mm}\\
	& & & C-A & I & 0.69 $\pm$ 0.14 & 4.8 & Microlensing \\ \cline{4-8} \vspace*{-3mm}\\
	& & & D-A & I & 0.90 $\pm$ 0.36 & 2.5 & Microlensing \\ \cline{4-8} \vspace*{-3mm}\\
	& & & C-B & I & 0.13 $\pm$ 0.06 & 2.2 & Microlensing \\ \cline{3-8} \vspace*{-3mm}\\
    & & Q 1017-207 & B-A & I & -0.28 $\pm$ 0.16 & 1.8 \\ \cline{3-8} \vspace*{-3mm}\\
    & & SDSS J1138+0314 & C-B & I & 0.13 $\pm$ 0.05 & 2.9 & Microlensing \\ \cline{3-8} \vspace*{-3mm}\\
    & & SDSS J1206+4332 & B-A & II & 0.31 $\pm$ 0.16 & 2.0 & Microlensing \\ \cline{3-8} \vspace*{-3mm}\\
    & & HE 2149-2745 & B-A & III & -0.15 $\pm$ 0.07 & 2.1 & Microlensing \\ \cline{2-8} \vspace*{-3mm}\\
	& \multirow{17}{*}{red wing} & \multirow{3}{*}{HE 0435-1223} & B-A & IV & 0.20 $\pm$  0.07 & 2.8 & Microlensing \\ \cline{4-8} \vspace*{-3mm}\\
	& & & D-B & IV & -0.12 $\pm$ 0.08 & 1.5 & \\ \cline{4-8} \vspace*{-3mm}\\
	& & & D-C & I & -0.39 $\pm$ 0.27 & 1.5 & \\ \cline{3-8} \vspace*{-3mm}\\
    & & HS 0818-1227 & B-A & I & -0.39 $\pm$ 0.19 & 2.0 & Microlensing \\ \cline{3-8} \vspace*{-3mm}\\
    & & SBS 0909+532 & B-A & I & 0.13 $\pm$ 0.07 & 1.9 & \\ \cline{3-8} \vspace*{-3mm}\\
    & & \multirow{2}{*}{Q 0957+561} & \multirow{2}{*}{B-A} & I & -0.19 $\pm$ 0.11 & 1.7 & \\
    & & & & III & -0.36 $\pm$ 0.17 & 2.2 & Microlensing \\ \cline{3-8} \vspace*{-3mm}\\
	& & \multirow{6}{*}{SDSS 1004+4112} & \multirow{4}{*}{B-A} & I & 0.30 $\pm$ 0.06 & 5.1 & Microlensing \\
	& & & & II & -0.53 $\pm$ 0.31 & 1.7 & \\
	& & & & III & -0.52 $\pm$ 0.15 & 3.6 & Microlensing \\ 
    & & & & IV & -0.65 $\pm$ 0.22 & 2.9 & Microlensing \\ \cline{4-8} \vspace*{-3mm}\\
	& & & C-B & I & 0.35 $\pm$ 0.10 & 3.5 & Microlensing \\ \cline{4-8} \vspace*{-3mm}\\
	& & & D-B & I & 0.40 $\pm$ 0.26 & 1.6 & \\ \cline{3-8} \vspace*{-3mm}\\
    & & \multirow{2}{*}{SDSS J1206+4332} & \multirow{2}{*}{B-A} & I & -0.27 $\pm$ 0.15 & 1.8 \\
    & & & & II & -0.46 $\pm$ 0.13 & 3.4 & Microlensing \\ \cline{3-8} \vspace*{-3mm}\\
	& & SDSS J1339+1310 & B-A & III & -0.62 $\pm$ 0.15 & 4.2 & Microlensing \\ \cline{3-8} \vspace*{-3mm}\\
    & & SBS 1520+530 & B-A & I & -0.25 $\pm$ 0.13 & 1.9 & \\ 
	\hline \vspace*{-3mm}\\
	\multirow{8}{*}{C III]} & \multirow{4}{*}{blue wing} & SDSS J0246-0825 & B-A & I & 0.13 $\pm$ 0.06 & 2.4 & Microlensing \\ \cline{3-8} \vspace*{-3mm}\\
    & & SDSS 1004+4112 & C-A & I & 0.33 $\pm$ 0.18 & 1.9 & \\ \cline{3-8} \vspace*{-3mm}\\
    & & SDSS J1155+6346 & B-A & I & -0.58 $\pm$ 0.39 & 1.5 & \\ \cline{3-8} \vspace*{-3mm}\\
    & & WFI 2033-4723 & C-B & II & 0.12 $\pm$ 0.06 & 2.1 & Microlensing \\ \cline{2-8} \vspace*{-3mm}\\
	& \multirow{4}{*}{red wing} & \multirow{2}{*}{HE 0435-1223} & B-A & I & 0.29 $\pm$  0.14 & 2.0 & Microlensing \\ \cline{4-8} \vspace*{-3mm}\\
	& & & C-A & I & 0.29 $\pm$ 0.15 & 1.9 & \\ \cline{3-8} \vspace*{-3mm}\\
	& & HE 1104-1805 & B-A & II & 0.21 $\pm$  0.14 & 1.5 & \\ \cline{3-8} \vspace*{-3mm}\\
    & & SDSS J1155+6346 & B-A & I & -0.63 $\pm$ 0.26 & 2.4 & Microlensing \\
	\hline \vspace*{-3mm}\\
	\multirow{3}{*}{Mg II} & \multirow{2}{*}{blue wing} &  HE 0435-1223 & D-B & II & 0.11 $\pm$  0.04 & 3.0 & Microlensing \\ \cline{3-8} \vspace*{-3mm}\\
    & & SDSS J1206+4332 & B-A & II & 0.23 $\pm$ 0.14 & 1.6 & \\ \cline{2-8} \vspace*{-3mm}\\
	& \multirow{1}{*}{red wing} & SDSS J1206+4332 & B-A & II & 0.28 $\pm$ 0.17 & 1.7 & \\ \hline
\end{tabular}\\
\flushleft *magnitude difference between pairs of images for the same epoch\\
**standard deviation of magnitude difference
\end{table*}

\begin{table*}
\setlength{\tabcolsep}{7.8pt}
\renewcommand{\arraystretch}{1}
\caption{Differences between epochs in the C IV, C III], and Mg II lines.}
\label{table3}
\centering
\begin{tabular}{lccccccc}
\hline\hline
Emission Line & Wing & Object & Image & Epoch & $\Delta$m$^{*}$ $\pm$ $\sigma^{**}$ & $\Delta$m/$\sigma$ & Classification\\
\hline \vspace{-3mm} \\ 
	\multirow{31}{*}{C IV} & \multirow{17}{*}{blue wing} & Q 0142-100 & A & II-I & -0.07 $\pm$ 0.05 & 1.5 & \\ \cline{3-8} \vspace*{-3mm}\\
    & & \multirow{7}{*}{SDSS 1004+4112} & \multirow{6}{*}{A} & II-I & 0.28 $\pm$ 0.07 & 4.1 & Microlensing Variability\\
    & & & & III-I & 0.18 $\pm$ 0.04 & 4.2 & Microlensing Variability\\
    & & & & IV-I & 0.43 $\pm$ 0.12 & 3.7 & Microlensing Variability\\
    & & & & III-II & -0.10 $\pm$ 0.06 & 1.5 & \\
    & & & & IV-II & 0.14 $\pm$ 0.09 & 1.6 & \\
    & & & & IV-III & 0.19 $\pm$ 0.08 & 2.2 & Microlensing Variability\\ \cline{4-8} \vspace*{-3mm}\\
    & & & B & IV-I & 0.20 $\pm$ 0.12 & 1.7 & \\ \cline{3-8} \vspace*{-3mm}\\
    & & \multirow{6}{*}{HE 1104-1805} & \multirow{3}{*}{A} & III-II & -0.28 $\pm$ 0.15 & 1.8 & \\
    & & & & IV-III & 0.14 $\pm$ 0.08 & 1.7 & \\
    & & & & V-III & 0.22 $\pm$ 0.08 & 2.9 & Intrinsic Variability? \\ \cline{4-8} \vspace*{-3mm}\\
    & & & \multirow{3}{*}{B} & III-I & -0.44 $\pm$ 0.23 & 1.9 & \\
    & & & & IV-I & -0.23 $\pm$ 0.12 & 2.0 & Intrinsic Variability?\\
    & & & & V-III & 0.29 $\pm$ 0.15 & 1.9 & \\ \cline{3-8} \vspace*{-3mm}\\
    & & SDSS J1339+1310 & A & III-II & -0.18 $\pm$ 0.07 & 2.4 & Microlensing Variability?\\ \cline{3-8} \vspace*{-3mm}\\
    & & \multirow{2}{*}{HE 2149-2745} & A & II-I & 0.12 $\pm$ 0.03 & 3.7 & Intrinsic Variability?\\ \cline{4-8} \vspace*{-3mm}\\
    & & & B & III-II & -0.20 $\pm$ 0.05 & 3.7 & Microlensing Variability \\ \cline{2-8} \vspace*{-3mm}\\
	& \multirow{14}{*}{red wing} & Q 0142-100 & A & II-I & -0.20 $\pm$ 0.08 & 2.5 & Intrinsic Variability?\\ \cline{3-8} \vspace*{-3mm}\\
    & & HE 0435-1223 & C & III-I & -0.49 $\pm$ 0.34 & 1.5 & \\ \cline{3-8} \vspace*{-3mm}\\
    & & SBS 0909+532 & A & II-I & -0.14 $\pm$ 0.06 & 2.2 & Microlensing Variability?\\ \cline{3-8} \vspace*{-3mm}\\
       & & \multirow{2}{*}{SDSS 1004+4112} & \multirow{2}{*}{A} & III-I & 0.26 $\pm$ 0.08 & 3.1 & Microlensing Variability\\
    & & & & IV-I & 0.52 $\pm$ 0.23 & 2.3 & Microlensing Variability\\ \cline{3-8} \vspace*{-3mm}\\
    & & \multirow{5}{*}{HE 1104-1805} & \multirow{3}{*}{A} & III-II & -0.24 $\pm$ 0.06 & 4.2 & Intrinsic Variability?\\
    & & & & IV-II & -0.15 $\pm$ 0.07 & 2.3 & Intrinsic Variability?\\
    & & & & IV-III & 0.09 $\pm$ 0.06 & 1.5 & \\ \cline{4-8} \vspace*{-3mm}\\
    & & & \multirow{2}{*}{B} & III-I & -0.14 $\pm$ 0.09 & 1.7 & \\
    & & & & IV-I & -0.10 $\pm$ 0.05 & 2.0 & Intrinsic Variability?\\ \cline{3-8} \vspace*{-3mm}\\
    & & \multirow{2}{*}{SDSS J1339+1310} & A & III-II & -0.29 $\pm$ 0.05 & 6.4 & Microlensing Variability\\ \cline{4-8} \vspace*{-3mm}\\
    & & & B & II-I & -0.16 $\pm$ 0.09 & 1.7 & \\ \cline{3-8} \vspace*{-3mm}\\
    & & \multirow{2}{*}{HE 2149-2745} & \multirow{2}{*}{A} & II-I & 0.23 $\pm$ 0.15 & 1.5 & \\
    & & & & III-II & -0.24 $\pm$ 0.14 & 1.7 & \\ \cline{1-8} \vspace*{-3mm}\\
 	\multirow{4}{*}{C III]} & \multirow{2}{*}{blue wing} & SDSS 1004+4112 & A & IV-II & -0.26 $\pm$ 0.16 & 1.7 & \\ \cline{3-8} \vspace*{-3mm}\\
    & & Q 0957+561 & B & III-I & -0.24 $\pm$ 0.16 & 1.5 & \\ \cline{2-8} \vspace*{-3mm}\\
	& \multirow{2}{*}{red wing} & \multirow{2}{*}{Q 0957+561} & A & III-I & -0.46 $\pm$ 0.25 & 2.5 & Intrinsic Variability?\\ \cline{4-8} \vspace*{-3mm}\\
    & & & B & III-I & -0.81 $\pm$ 0.53 & 1.5 & \\ \cline{1-8} \vspace*{-3mm}\\
    \multirow{8}{*}{Mg II} & \multirow{6}{*}{blue wing} & \multirow{3}{*}{HE 0435-1223} & A & IV-III & -0.12 $\pm$ 0.08 & 1.5 & \\ \cline{4-8} \vspace*{-3mm}\\
    & & & \multirow{2}{*}{D} & III-II & -0.21 $\pm$ 0.11 & 1.9 & \\
    & & & & IV-III & -0.20 $\pm$ 0.12 & 1.6 & \\ \cline{3-8} \vspace*{-3mm}\\
    & & \multirow{2}{*}{FBQ 0951+2635} & \multirow{2}{*}{A} & II-I & -0.11 $\pm$ 0.07 & 1.5 & \\
    & & & & III-I & -0.15 $\pm$ 0.09 & 1.7 & \\ \cline{3-8} \vspace*{-3mm}\\
    & & WFI 2033-4723 & C & III-II & 0.12 $\pm$ 0.07 & 1.6 & \\ \cline{2-8} \vspace*{-3mm}\\
	& \multirow{2}{*}{red wing} & HE 0435-1223 & D & IV-III & -0.37 $\pm$ 0.23 & 1.6 & \\ \cline{3-8} \vspace*{-3mm}\\
    & & SDSS J1353+1138 & B & II-I & 0.33 $\pm$ 0.20 & 1.6 & \\ \hline
\end{tabular}
\flushleft *magnitude difference between epochs for the same image\\
**standard deviation of magnitude difference
\end{table*}

\section{Results}\label{sec3}
In general, there is a good match between the emission line profiles of different images at different epochs, as seen in Figures \ref{em1} to \ref{em4}. The Mg II line is only weakly affected by either microlensing or intrinsic variability (consistent with the results presented in \citealt{Fian2018}). Only one system (HE 0435-1223) shows a small change (0.11 mag) with S/N > 2 in the selected integration window for the blue wing as defined in Figure \ref{em1}. Visually inspecting the data, we find changes in the extreme red wings of SBS 0909+532 and SDSS J1206+0118. These changes could have their origin in the UV Fe II blend $\lambda\lambda$2861-2917 located redward of Mg II, showing up with different intensities for different quasar spectra (see \citealt{Kova2015}). \\

\begin{figure*}
\centering
\includegraphics[width=\textwidth]{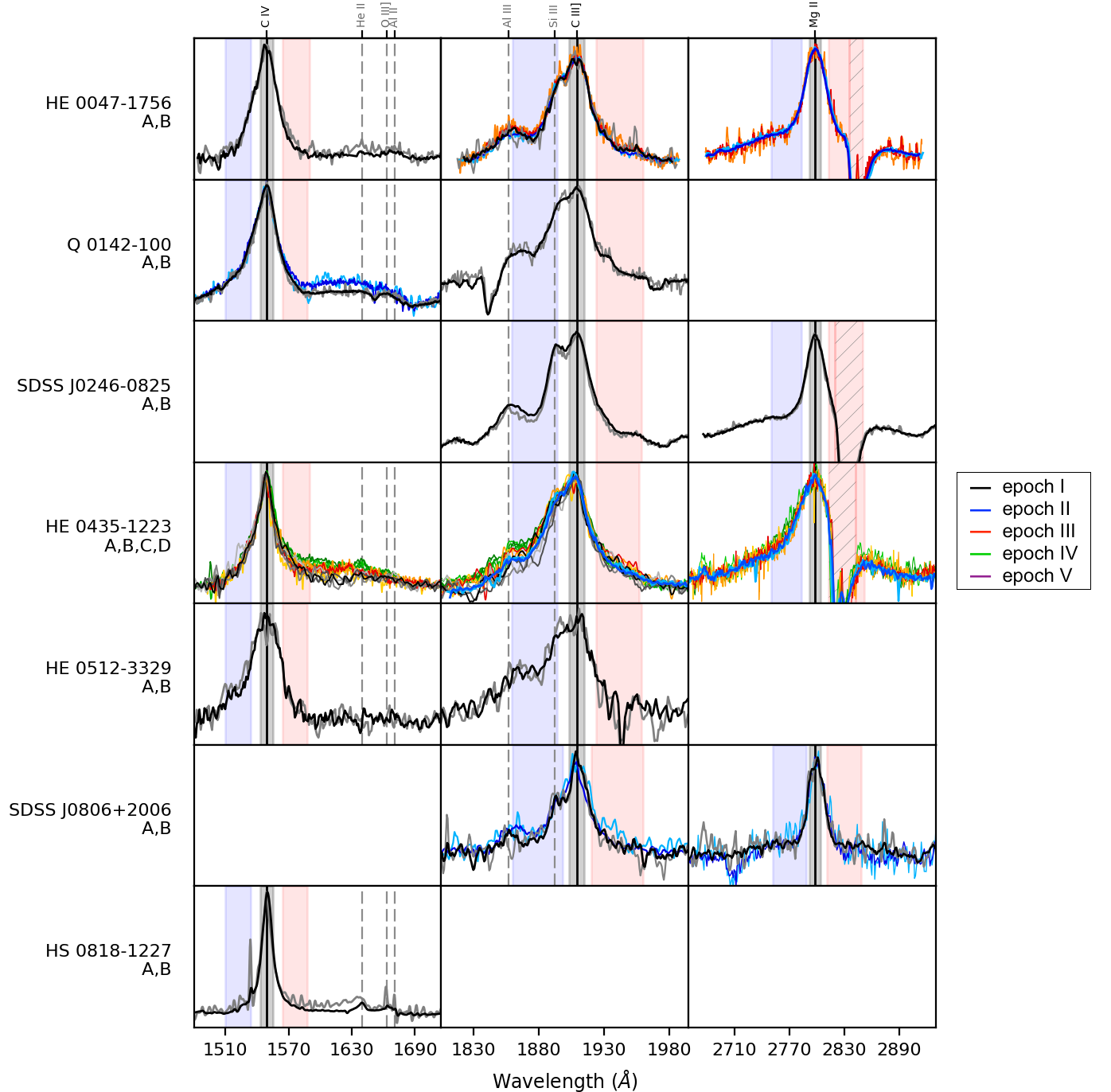}
\caption{C IV, C III], and Mg II emission line profiles from different epochs and images overimposed after subtracting the continuum and matching the line core. Different color shadings stand for different images in the corresponding epoch. Blue and red shaded regions show the integration windows used for the magnitude difference calculations. The flux is in arbitrary units.}
\label{em1}
\end{figure*}

\begin{figure*}
\centering
\includegraphics[width=\textwidth]{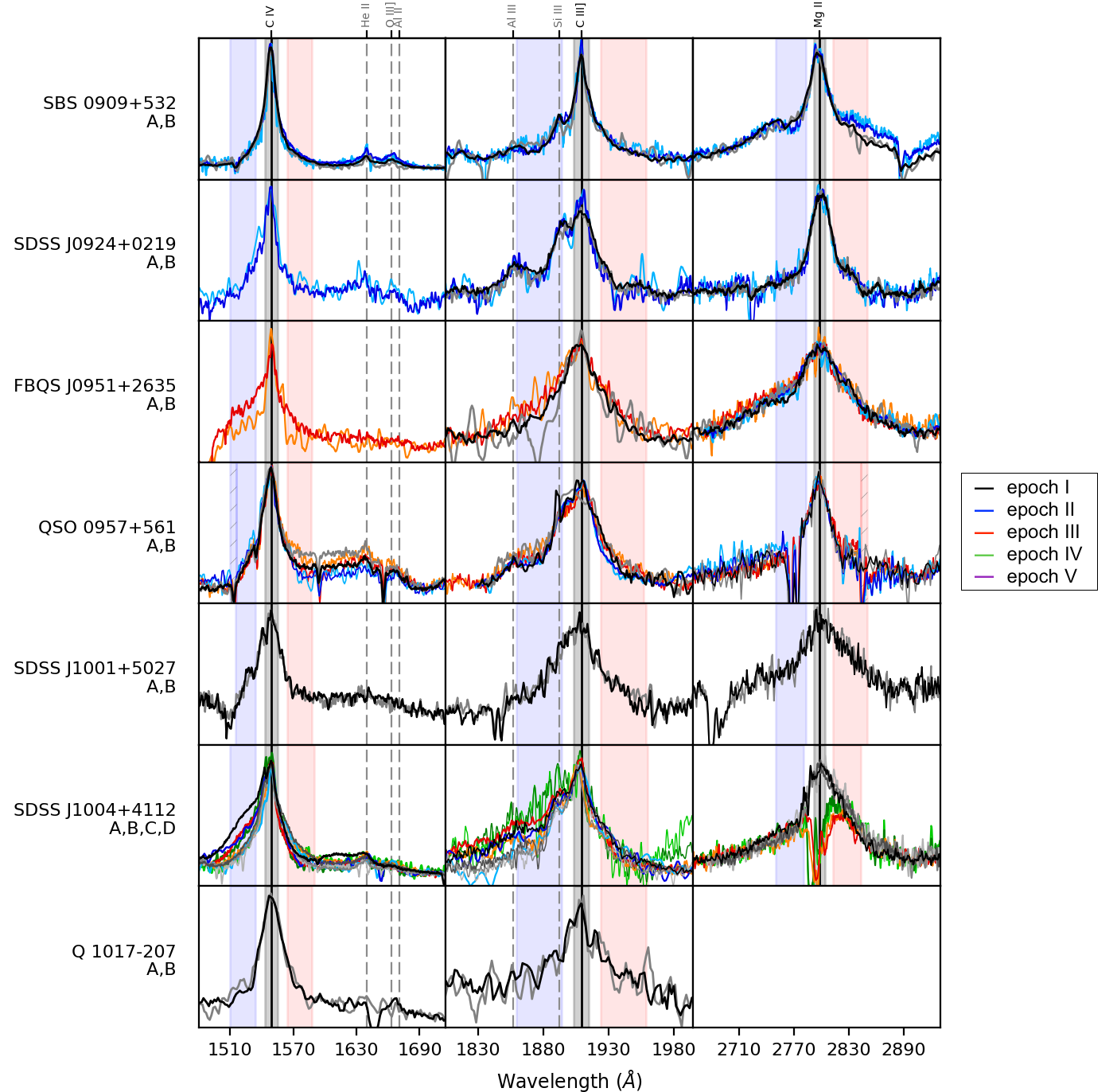}
\caption{Continuation of Figure \ref{em1}.}
\label{em2}
\end{figure*}

\begin{figure*}
\centering
\includegraphics[width=\textwidth]{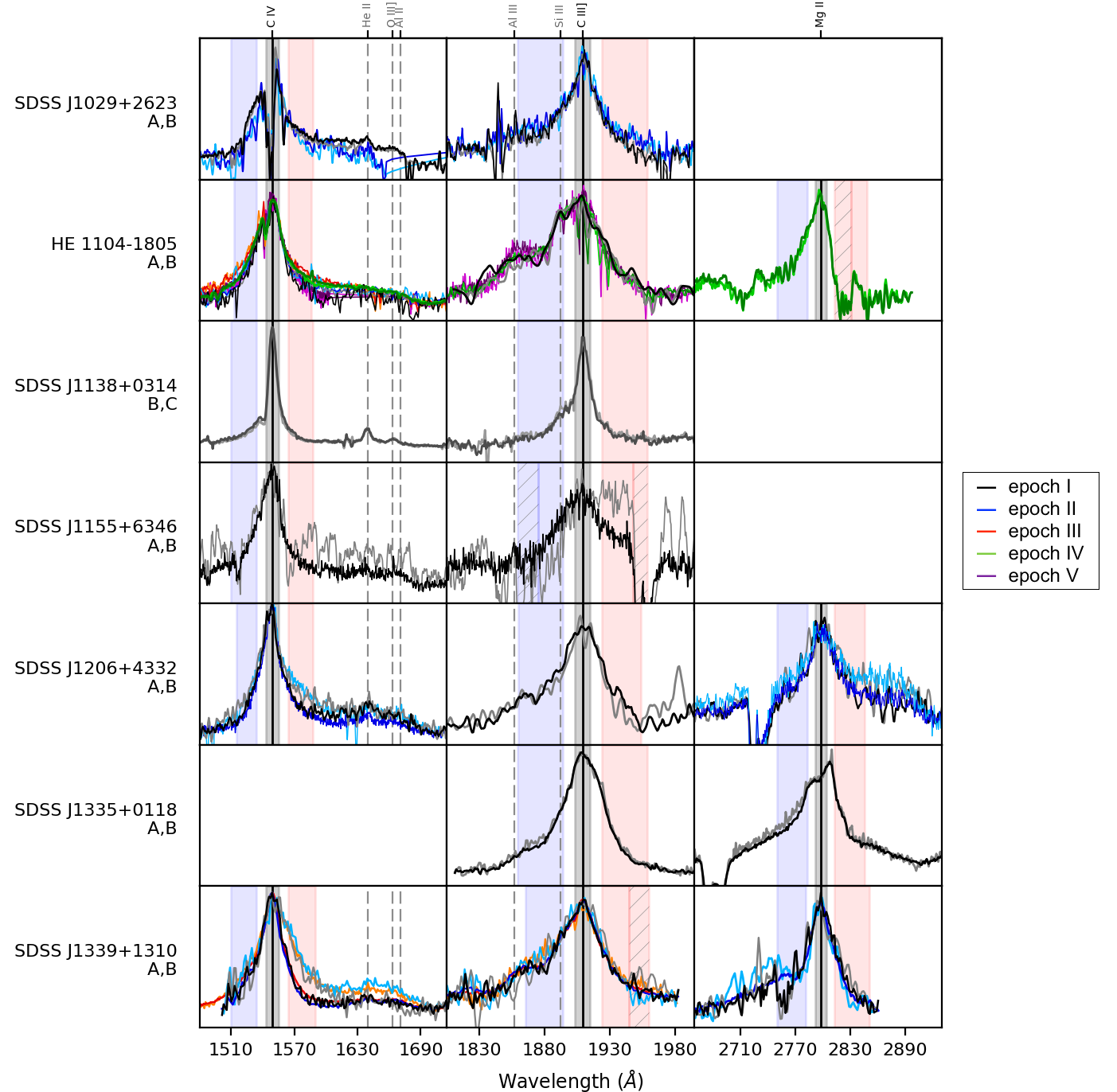}
\caption{Continuation of Figure \ref{em1}.}
\label{em3}
\end{figure*}

\begin{figure*}
\centering
\includegraphics[width=\textwidth]{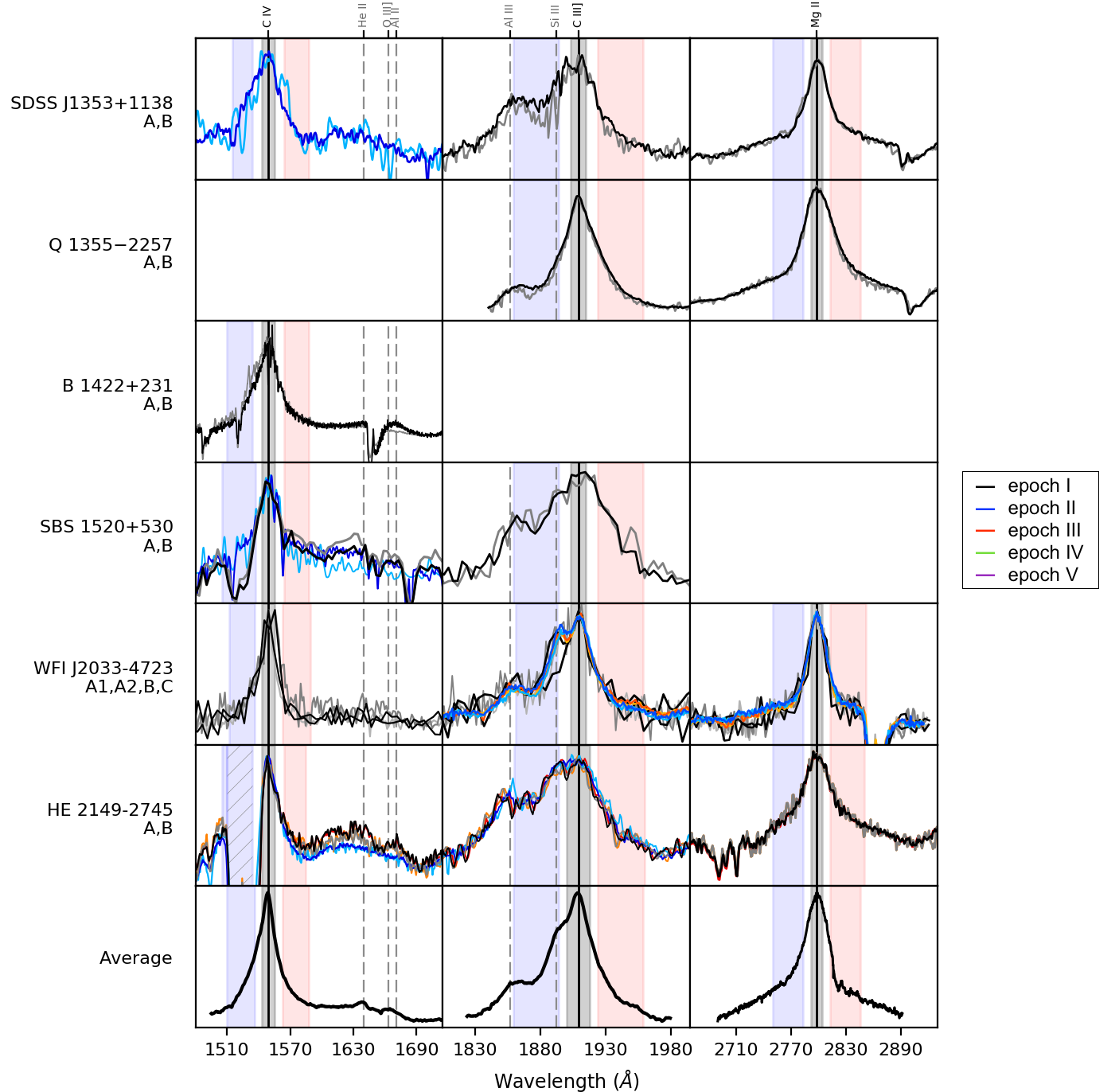}
\caption{Continuation of Figure \ref{em1}.}
\label{em4}
\end{figure*}

At odds with the results obtained from the limited sample presented in Paper I, we find, on the basis of more available epochs, that the CIII] line seems to be significantly more affected by microlensing than the Mg II line, indicating that it arises from a more compact region inside the BLR. We detect microlensing at the 2$\sigma$ level within the defined integration windows in four systems (SDSS J0246-0825, HE 0435-1223, SDSS J1155+6346, WFI 2033-4723), whereas evidence of intrinsic variability (at the 2$\sigma$ level) can only be found in one system (Q0957+561). Also, the systems SDSS 1004-4112, SDSS J1155+6346, and HE 1104-1805 show microlensing differences at a smaller S/N ($\sigma$ $\geq$ 1.5). Notice the impact 
on the statistics of the quadruple systems HE 0435-1223 and SDSS J1004+4112, which have 
been observed in four epochs.\\

In the case of C IV there are obvious differences in the selected windows of the line profiles with nearly half of the sample showing changes due to microlensing and/or intrinsic variability. After consistently separating microlensing from intrinsic variability according to the criteria explained in Section \ref{sec2.2} (see also Tables \ref{table2} and \ref{table3}), we find that nine objects are clearly affected by microlensing at the 2$\sigma$ level (HE 0435-1223, HS 0818-1227, FBQS J0951+2635, QSO 0957+561, SDSS J1004+4112, SDSS J1138+0314, SDSS J1206+4332, SDSS J1339+1310, HE 2149-2745) and two systems show evidence of intrinsic variability at the 2$\sigma$ level (HE 1104-1805, HE 2149-2745). The fact that for some systems C IV does not show microlensing anomalies may be because its emission region is either too big or it lies in a region without significant microlensing fluctuations. However, the non-detection of microlensing also puts constrains on the size of the emission region and should be taken into account. Additional three systems (SBS 0909+532, Q1017-207, SBS 1520+530) show evidence of microlensing at the 1.5$\sigma$ level, and Q 0142-100 seems to be dominated by intrinsic variability. The differences observed in QSO 0957+561 may be explained by intrinsic variability combined with the large time-delay between the images of this double plus a possible contribution from microlensing. It is important to notice that intrinsic variability mainly affects the wings of C IV, but, in contrast with microlensing, it does not induce marked asymmetries in the line profile.\\

In Figure \ref{ms1} and \ref{ms2} we show, for systems with at least two epochs of observation, the average spectra with one and two sigma intervals to emphasize the variability in the wings. There are several systems (Q 0142-100, HE 0435-1223, FBQS J0951+2635, QSO 0957+561, SDSS J1004+4112, HE 1104-1805, SDSS J1206+4332, SDSS J1339+1310, HE 2149-2745) in which microlensing and/or intrinsic variability is strongly affecting the wings of C IV (mainly the red wing) with little or no traces of variability in the wings of C III] and Mg II (except the blue wing of C III] in SDSS J1004+4112 and the extreme red wing of Mg II in SDSS J1206+4332).\\

\begin{figure*}
\centering
\includegraphics[width=\textwidth]{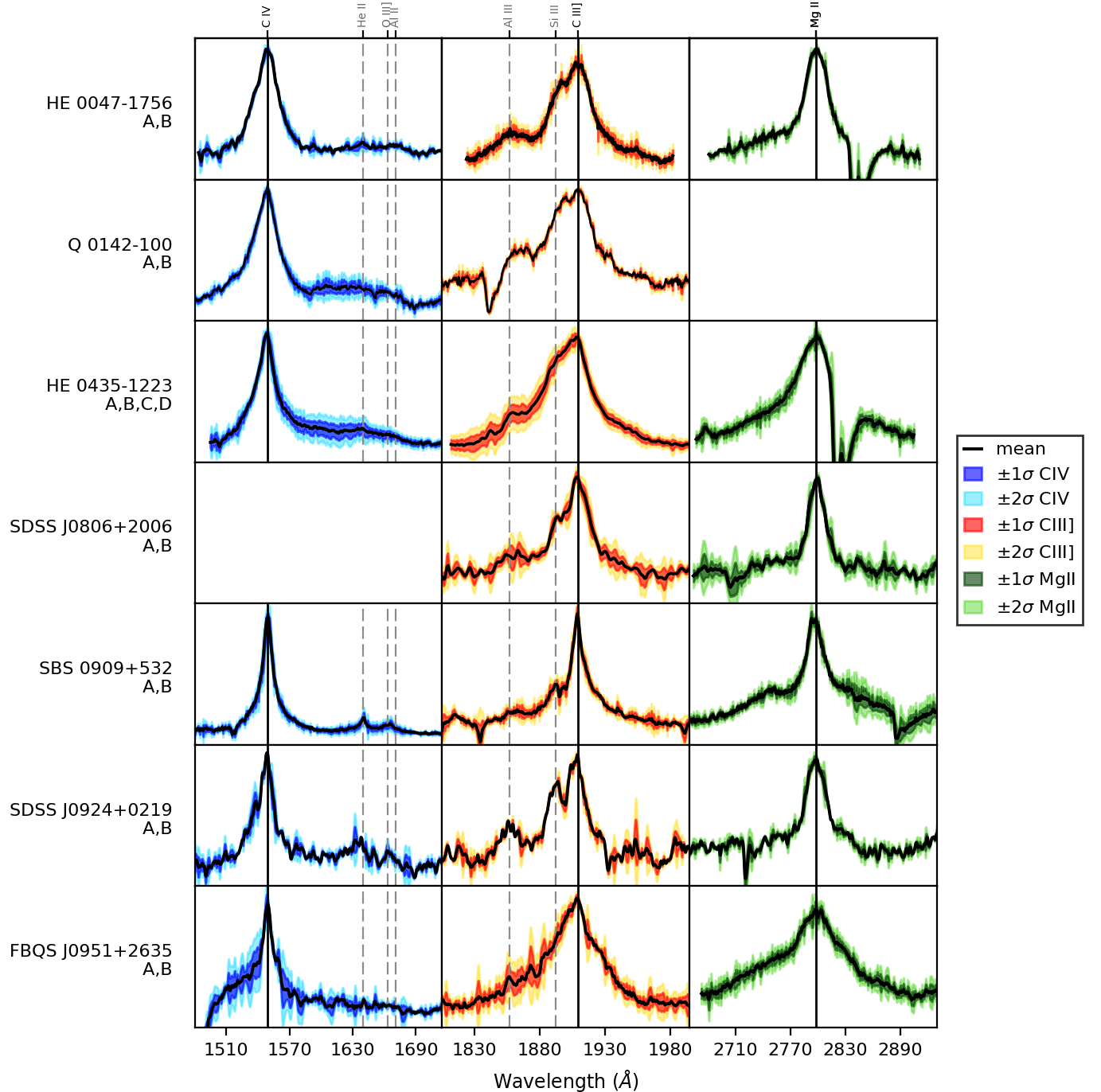}
\caption{Average (black) spectra for the C IV, C III], and Mg II emission lines with one (dark blue, red, and green) and two (light blue, red, and green) sigma intervals. We did not take into account noisy spectra and spectra that show absorption features in the line core (as in the case of Mg II in SDSS J1004+4112). The y-axis is in arbitrary units of flux.}
\label{ms1}
\end{figure*}

\begin{figure*}
\centering
\includegraphics[width=\textwidth]{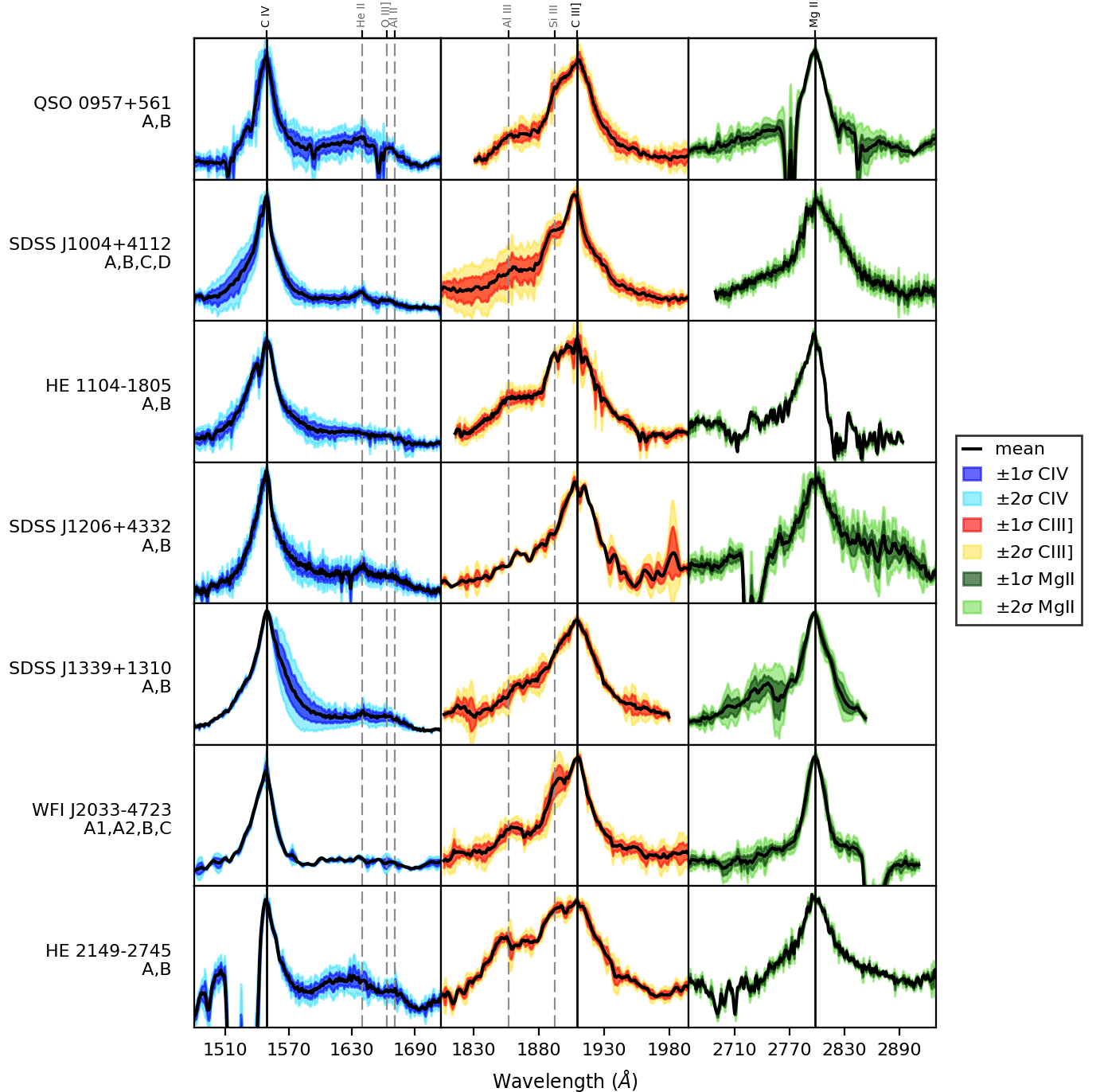}
\caption{Continuation of Figure \ref{ms2}.}
\label{ms2}
\end{figure*}

\begin{figure}
\centering
\includegraphics[width=0.5\textwidth]{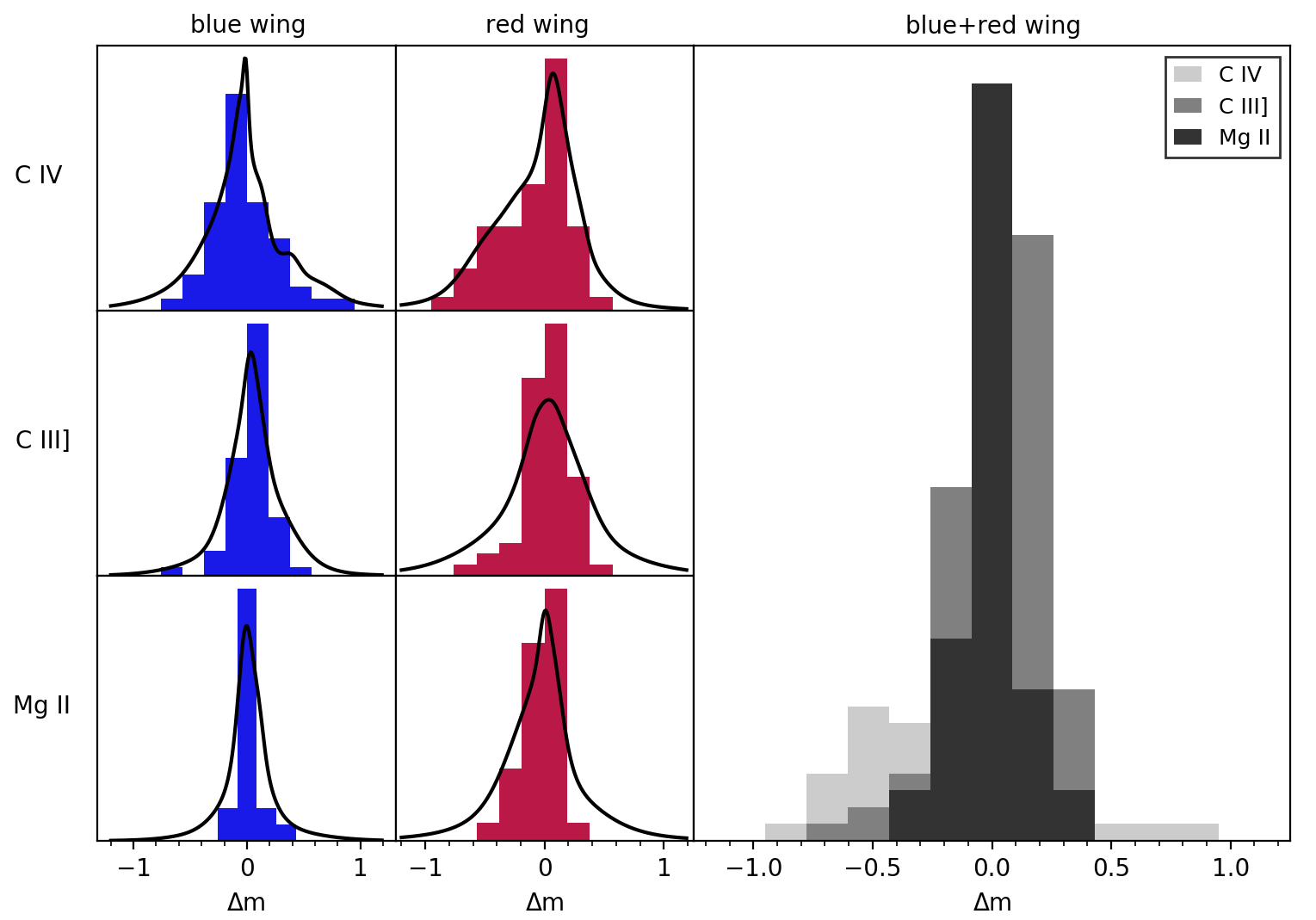}
\caption{Histograms of microlensing magnification difference between image pairs (in the same epoch) in the blue and red wings of C IV, C III], and Mg II. Black curves show the corresponding Gaussian kernel density estimates of the PDFs.}
\label{histo_images}
\end{figure}

\begin{figure}
\centering
\includegraphics[width=0.5\textwidth]{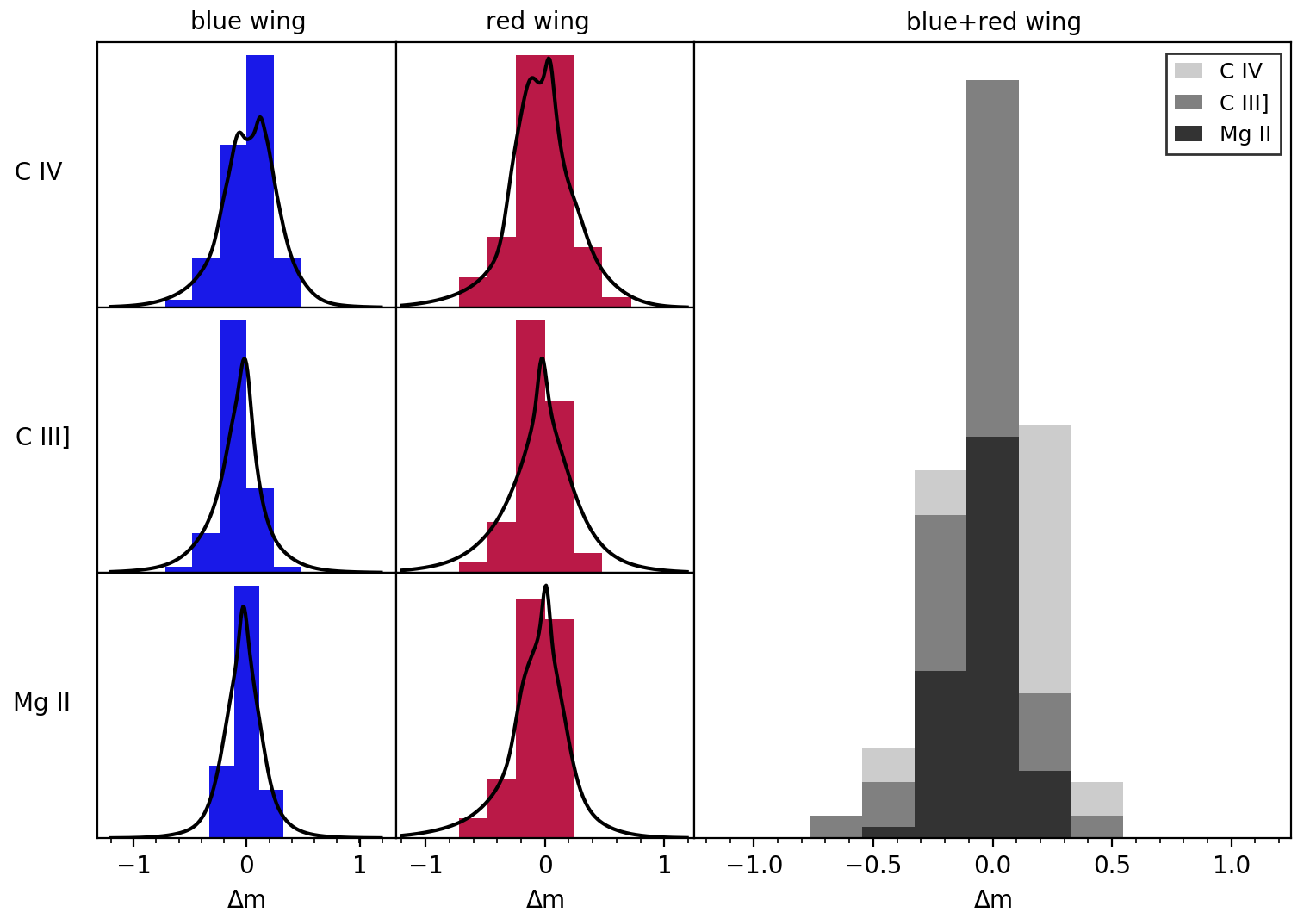}
\caption{Histograms of magnitude difference for the same image at different epochs in the blue and red wings of C IV, C III], and Mg II. Black curves show the corresponding Gaussian kernel density estimates of the PDFs.}
\label{histo_epochs}
\end{figure}

In Figures \ref{histo_images} and \ref{histo_epochs} we present histograms of the differences between images and epochs for the C IV, C III], and Mg II lines. We have overlaid the corresponding Gaussian kernel density estimates of the probability density functions (PDFs) to consider the impact of errors in the individual measurements. The first thing to note is that microlensing and intrinsic variability effects diminish according to the sequence: C IV > C III] > Mg II. The second thing to note is that the differences between the images are more or less on the same order of magnitude as differences between the epochs.

%%%%%%%%%%%%%%%%%%%%%%%%%%%%%%%%%%%%%%%%%%%%%%%%%%%%%%%%%%%%%%%%%%%%%%%%%%%%%%%%%%%%%%%%%%%%
%%%%%%%%%%%%%%%%%%%%%%%%%%%%%%%%%%%%%%%%%%%%%%%%%%%%%%%%%%%%%%%%%%%%%%%%%%%%%%%%%%%%%%%%%%%%
\section{Structure of the broad line region}\label{sec4} 
Given the estimates of the differential microlensing in the wings between pairs of images in each system for different emission lines, we can estimate the size of their emission region and thus reveal the structure of the BLR in lensed quasars. We treat each microlensing measurement as a single epoch event and from the microlensing magnification corresponding to all the image pairs, in all available epochs of observation, we compute the joint microlensing probability, $P(r_s)$, to obtain an average estimate of the size. We follow the steps described in \citet{Guerras2013},
\begin{eqnarray}
P(r_s)=\prod\limits_{i} P_i(r_s), \\
P_i(r_s)\propto e^{-\frac{{\chi_i^2(r_s)}}{2}},\\
\chi_i^2(r_s)=\sum\limits_{\alpha_i}{\sum\limits_{\beta_i < \alpha_i}\left(\frac{{\Delta m^{obs}_{\beta_i\alpha_i}-\Delta m_{\beta_i\alpha_i}(r_s)}}{\sigma_{\beta_i\alpha_i}}\right)^2},
\end{eqnarray}
where $\Delta m^{obs}_{\beta_i\alpha_i}$ is the observed differential microlensing magnification between images $\alpha$ and $\beta$ of system $i$ and $\Delta m_{\beta_i\alpha_i}(r_s)$ is the differential microlensing magnification predicted by the simulations for a given value of $r_s$.\\

Our simulations are based in $2000\times2000$ pixel microlensing magnification maps, generated at the positions of the images using the Inverse Polygon Mapping method described in \citet{Mediavilla2006,Mediavilla2011}. To compute the magnification maps we used the local convergence $\kappa$ and the local shear $\gamma$, obtained by fitting a singular isothermal sphere with an external shear\footnote{In the case of Q 0957+561 we tested the robustness of our results concerning the macro-model by comparing it with the lens parameters from \citet{Pelt1998}, obtaining a slightly smaller value for the half-light radius of the BLR for this source after computing the magnification maps for images A and B and convolving them with Gaussian profiles of different sizes.} (SIS+$\gamma_e$) that reproduce the coordinates of the images (\citealt{Mediavilla2006}). The produced maps span $400\times400$ light-days$^2$ on the source plane, with a pixel size of 0.2 light-days. We assume a mean stellar mass of $M=0.3M_{\odot}$ and for the fraction of mass in stars we use $\alpha=0.1$. All linear sizes are rescaled with the square root of the microlens mass $\sqrt{M/M_\odot}$. To simulate the effect of finite sources we model the luminosity profile of the region emitting the wings as a Gaussian ($I\propto exp(-R^2/2r_s^2)$) and the magnifications experienced by a source of size $r_s$ are then found by convolving the magnification maps with the Gaussian profiles of sigma $r_s$. We used a logarithmic grid for the source sizes, spanning an interval between $\sim$1 to 100 light-days. These sizes can be converted to half-light radii multiplying by 1.18, $R_{1/2}=1.18r_s$.\\

\begin{figure*}
\centering
\includegraphics[width=0.95\textwidth]{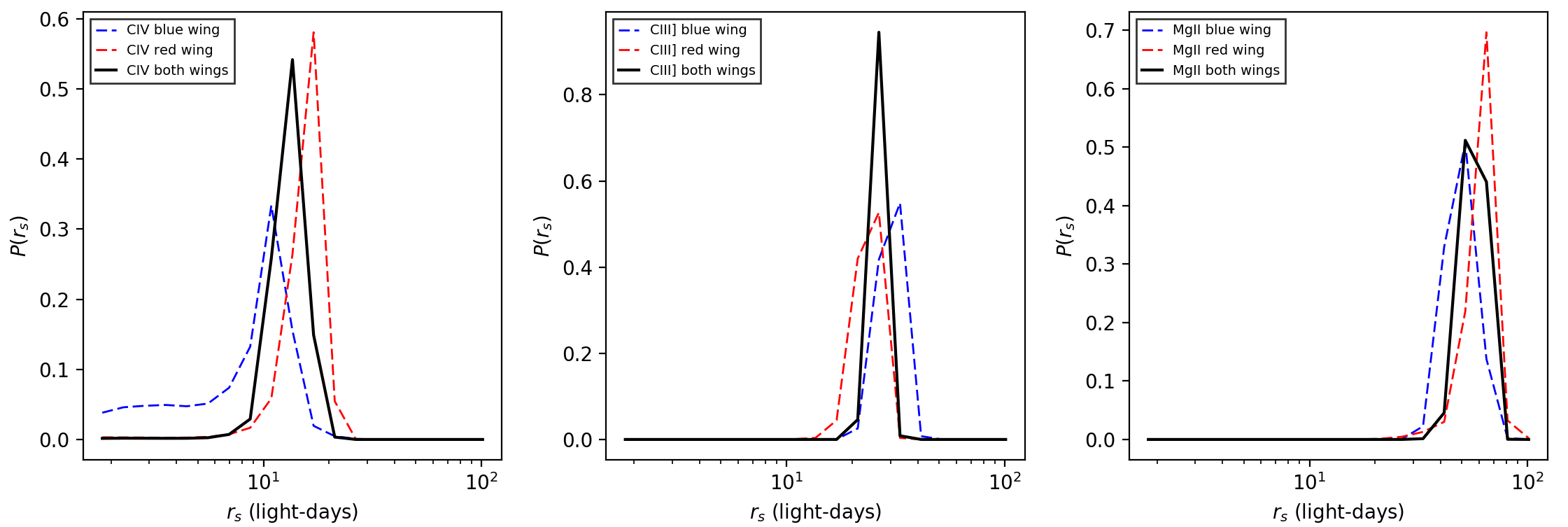}
\caption{Joint likelihood (solid black line) for C IV (left panel), C III] (middle panel), and Mg II (right panel). Dashed lines show the Joint likelihood functions for the red and blue wing, respectively.}
\label{PDF}
\end{figure*}

The resulting joint likelihood functions, scaled to a mean microlens mass of $\langle M \rangle = 1 M_\odot$, for the C IV, C III], and Mg II emission lines can be seen in Figure \ref{PDF}. From the right panel in Figure \ref{PDF} we can 
infer (using a logarithmic prior for the size) a size of $r_s = 57.0_{-13.3}^{+3.2} \sqrt{M/M_{\odot}}$ light-days ($68\%$ confidence) for the region emitting the Mg II line. This result is in good agreement with the size of the LIL obtained by \citet{Guerras2013} ($r_s = 55_{-35}^{+150} \sqrt{M/M_{\odot}}$ light-days). We note that the uncertainties in our calculation are much smaller than the large errors found by \citet{Guerras2013}. From the likelihood function (middle panel of Figure \ref{PDF}) corresponding to the observed microlensing we estimate a size of $r_s = 26.3_{-3.4}^{+1.6} \sqrt{M/M_{\odot}}$ light-days for the region emitting C III]. As commented in Section \ref{sec3}, this relatively small size is dominated by the impact of two quadruples, HE 0435-1223 and SDSS J1004+4112, which have been observed in four epochs. If we exclude these systems (and the noisy data from SDSS J1155+6346), we obtain a size of around 40 light-days. We note that this line is blended with Al III and Si III in the wavelength range used to define the blue wing; therefore, we are likely underestimating the size of its emitting region. For C IV, we obtain a size of $r_s = 13.1_{-3.3}^{+0.7} \sqrt{M/M_{\odot}}$  light-days, indicating that part of this line is formed in the inner part of the BLR close to the accretion disk. The strong impact of microlensing in the wings of C IV may be partially related to the red-shelf at around 1610\AA\ and the complex formed by the He II, O III], and Al II lines. The presence of these emission lines as well as the subjacent iron pseudo-continuum could tend to bias the size estimate toward smaller values.

\subsection{Core microlensing}
\begin{figure}
\centering
\includegraphics[width=0.5\textwidth]{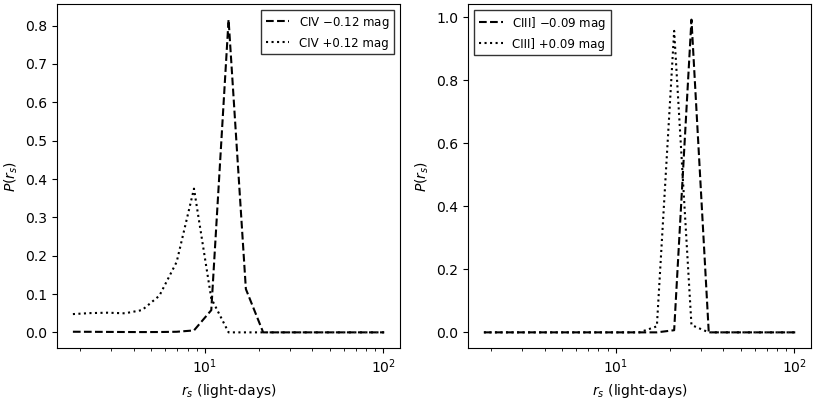}
\caption{Joint likelihood for C IV (left panel) and C III] (right panel) after subtracting (dashed lines) and adding (dotted lines) the average impact of microlensing of the corresponding core.}
\label{cores}
\end{figure}

Emission line cores of low-ionization lines are supposed to be insensitive to microlensing as they arise from spatially extended regions, large enough to smear out the effects of microlensing. However, the cores of high-ionization lines might be affected by microlensing, thereby biasing the estimated BLR size toward smaller values. In \citet{Fian2018} we measured the impact of microlensing on the C IV and C III] line cores, finding an average difference of $0.12\pm0.11$ mag for C IV (68\% confidence interval) and $0.09\pm0.08$ mag for C III]. To evaluate the impact of core microlensing on our BLR size estimates, we added or subtracted these mean differences to or from each microlensing measurement. While subtracting the corresponding difference from the microlensing measurements only induces small changes in the size of less than 1 light-day, adding them leads to (as expected) slightly smaller sizes ($\Delta r_s \sim6$ light-days for C IV, and $\Delta r_s \sim5$ light-days for C III]). Figure \ref{cores} and Table \ref{core} summarize the results.

\begin{table}[h]
%\centering
%\tabcolsep=0.7cm
\renewcommand{\arraystretch}{1.4}
\caption{Scale radius $r_{s}$ (in light-days) of C IV and C III] inferred from Figure \ref{cores}, together with the original values inferred from Figure \ref{PDF}.}
\begin{tabu}to 0.49\textwidth {X[l]X[c]X[c]X[c]}
\hline
\hline
Line & $-ML_{ core}^{*}$ & Original & $+ML_{ core}^{**}$\\
\hline 
C IV & $13.7_{-1.0}^{+2.0}$ & $13.1_{-3.3}^{+0.7}$ & $6.8_{-3.2}^{+1.9}$ \\
C III] & $26.5_{-0.5}^{+0.5}$ & $26.3_{-3.4}^{+1.6}$ & $21.3_{-1.5}^{+2.1}$ \\
\hline 
\end{tabu}
\label{core}
\vspace*{-3mm}
\flushleft *corresponds to the dashed lines in Figure \ref{cores}\\
**corresponds to the dotted lines in Figure \ref{cores}
%\tablenotetext{*}{corresponds to the dashed lines in Figure \ref{cores}}\\
%\tablenotetext{**}{corresponds to the dotted lines in Figure \ref{cores}}
\end{table}%\newpage

\subsection{Intrinsic variability}
%\textbf{It is well known that quasars show temporal flux variabilities and that the color becomes bluer as the flux becomes brighter (\citealt{Wilhite2005,Cristiani1997}), meaning that the color of quasar changes with time (\citealt{Yonehara2008}). Due to the arrival time delay between the images, multiple images observed at the same epoch correspond to images at intrinsically different epochs. Hence, multiple images observed at the same time may show different colors. To evaluate this effect, we use Eq. (3) in \citet{Yonehara2008} to estimate the color change in a given time interval and for a given wavelength. We roughly estimate a median time delay of 47 days in the observer-frame for our sample of quasars (using values from the literature). With an average redshift of $z\sim2$, this converts to 16 days in the rest-frame. Adopting 16 days for the time scale, 1549\AA\ (or 1909\AA) for the rest-frame wavelength, and an average \textit{i}-band magnitude of $\sim19.5$ mag (estimated from the CASTLES Survey\footnote{\textbf{https://lweb.cfa.harvard.edu/castles/}} for our sample), we obtain the expected average flux variation of our sample of quasars at the wavelength of C IV (and C III]). Since the observed magnitude differences between images in the wings of CIV and CIII] are much higher than the expected change of $\sim0.12$ mag, it is impossible to reproduce all the observed differences only by this scenario. Hence, for most quasars in our sample intrinsic variability represents only a modest contribution to the measured magnitude difference.}\\
To check whether systems that show strong intrinsic variability in our sample have an impact on our average size estimates, we excluded individual systems where we found possible signatures of source variability (at the 2$\sigma$ level; see Table \ref{table3}) from our size analysis. In the case of C IV we have possibly detected prominent intrinsic variability for three systems (Q 0142-100, HE 1104-1805, and HE 2149-2745). Removing the affected epochs for these systems from our analysis leads to an average size of $r_s = 9.9_{-4.3}^{+3.7} \sqrt{M/M_\odot}$ light-days for the region emitting the C IV line ($\sim 20\%$ smaller than the original size estimate). In the case of C III] only one system (Q 0957+561) shows significant variability between two epochs of observation. Excluding them from our analysis results in an average size of $r_s = 28.3_{-1.8}^{+4.8} \sqrt{M/M_\odot}$ light-days ($\sim10\%$ bigger than the original value). The Mg II emission line does not show any evidence of intrinsic variability at the 2$\sigma$ level.

\subsection{Pure microlensing signal}\label{pm}

\begin{figure*}
\centering
\includegraphics[width=0.9\textwidth]{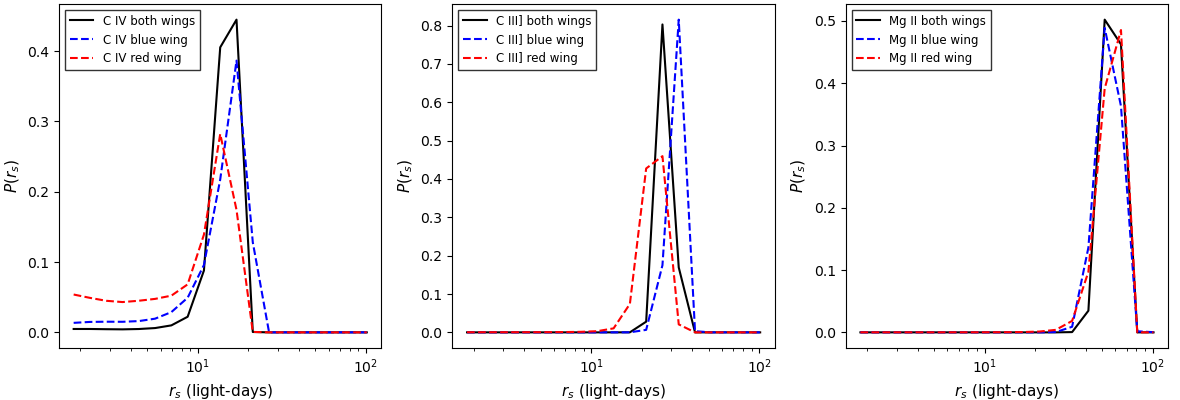}
\caption{Joint likelihood (solid black line) for C IV (left panel), C III] (middle panel), and Mg II (right panel) for "pure" microlensing signals (see Section \ref{pm}). Dashed lines show the joint likelihood functions for the red and blue wing, respectively.}
\label{pure_ML}
\end{figure*}

To obtain pure microlensing signals (i.e., $A(t_1)/B(t_1+\Delta t_{AB})$, where $\Delta t_{AB}$ is the time delay between the images A and B), we performed an additional analysis based on pairs of images at two epochs ($t_1$ and $t_2$) separated by times close or coincident with time delays between their images ($t_2 - t_1 \sim \Delta t_{AB}$). When time delays are short (in the range of a few days or weeks), the use of $A(t_1)/B(t_1)$ can also be regarded as "true" microlensing signal. In Table \ref{timedelays} we list all eligible systems (i.e., systems with known time delay $\leq 60$  days and systems where the separation between epochs equals approximately the time delay between the images; in total 13 out of 27 systems fulfill these criteria) together with the time delay between their images and (in the case of Q 0957+561) the temporal separation between their epochs of observation. In Figure \ref{pure_ML} we show the resulting joint likelihood functions for the C IV, CIII], and Mg II emission lines and we list the inferred sizes in Table \ref{sizes_dt}. We find that excluding systems with time delays $> 60$ days do not induce significant changes in the size estimates (see Table \ref{sizes_dt}). Hence, the microlensing effects are little contaminated by intrinsic variability modulated by the lens time delays.\\

\begin{table}
%\centering
\tabcolsep=0cm
\renewcommand{\arraystretch}{1.45}
\caption{Time delays between lensed images for systems with "pure" microlensing measurements.}
\begin{tabu}to 0.49\textwidth {X[l]X[l]X[l]}
\hline
\hline
Objects & Time Delay (days) & Reference\\
\hline 
HE 0047-1756 & $\Delta t_{AB} = -10.8_{-1.0}^{+1.0} $ & \citealt{Millon2020} \\ \hline 
SDSS J2046-0825 & $\Delta t_{AB} = 0.8_{-5.2}^{+5.0} $ &  \citealt{Millon2020XIX} \\ \hline
HE 0435-1223 &  $\Delta t_{AB} = -9.0_{-0.8}^{+0.8}$ & \citealt{Millon2020XIX} \\
& $\Delta t_{AC} = -0.8_{-0.7}^{+0.8} $ &  \\
& $\Delta t_{AD} = -13.8_{-0.8}^{+0.8} $ &  \\ \hline 
SBS 0909+532 & $\Delta t_{AB} = -50.0_{-4.0}^{+2.0}$ & \citealt{Hainline2013} \\ \hline 
SDSS J0924+0219 & $\Delta t_{AB} = +2.4_{-3.8}^{+3.8}$ & \citealt{Millon2020XIX} \\ \hline 
FBQ 0951+2635 & $\Delta_{AB} = +16.0_{-2.0}^{+2.0} $ & \citealt{Jakobsson2005} \\ \hline 
Q 0957+561 & $\Delta t_{AB} = -417_{-2.0}^{+2.0} $ & \citealt{Shalyapin2008}\\ 
& ($\Delta t_{12} = +414 $) & \\ \hline 
SDSS J1004+4112 & $\Delta_{AB} = -40.6_{-1.8}^{+1.8}$ &\vspace*{-3mm} \citealt{Fohlmeister2008} \vspace*{1mm}\\ \hline 
SDSS J1335+0118 & $\Delta_{AB} = -56.0_{-6.1}^{+5.7}$ & \citealt{Millon2020XIX} \\ \hline 
SDSS J1339+1310 & $\Delta_{AB} = +47.0_{-6.0}^{+5.0}$ & \vspace*{-3mm}\citealt{Goicoechea2016} \vspace*{1mm} \\ \hline 
B 1422+231 & $\Delta_{AB} = -1.5_{-1.4}^{+1.4} $ & \vspace*{-3mm} \citealt{Patnaik2001} \vspace*{1mm} \\ \hline 
WFI 2033-4723 & $\Delta t_{A_1B}= +36.2_{-1.6}^{+2.3}$ & \citealt{Bonvin2019} \\ 
& $\Delta t_{A_2B} = +37.3_{-2.6}^{+3.0} $&  \\
& $\Delta t_{BC} = -59.4_{-3.4}^{+1.3}$ &  \\ \hline
HE 2149-2745 & $\Delta t_{AB} = -39.0_{-16.7}^{+14.4}$ & \citealt{Millon2020XIX} \\ \hline
\hline 
\end{tabu}
\label{timedelays}
\end{table}

\begin{table}
%\centering
%\tabcolsep=0.7cm
\renewcommand{\arraystretch}{1.45}
\caption{Scale radii $r_{s}$ (in light-days) for C IV, C III], and Mg II inferred from Figure \ref{pure_ML}, together with the original values inferred from Figure \ref{PDF}.}
\begin{tabu}to 0.49\textwidth {X[l]X[c]X[c]X[c]}
\hline
\hline
Line & Original & Pure ML & Difference \\
\hline 
C IV & $13.1_{-3.3}^{+0.7}$ & $14.4_{-0.8}^{+2.6}$ & 1.3 \\
C III] & $26.3_{-3.4}^{+1.6}$ & $27.6_{-1.1}^{+5.5}$ & 1.3 \\
Mg II & $57.0_{-13.3}^{+3.2}$ & $56.2_{-4.4}^{+8.5}$ & 0.8 \\
\hline 
\end{tabu}
\label{sizes_dt}
\end{table}

%%%%%%%%%%%%%%%%%%%%%%%%%%%%%%%%%%%%%%%%%%%%%%%%%%%%%%%%%%%%%%%%%%%%%%%%%%%%%%%%%%%%%%%%%%%%
%%%%%%%%%%%%%%%%%%%%%%%%%%%%%%%%%%%%%%%%%%%%%%%%%%%%%%%%%%%%%%%%%%%%%%%%%%%%%%%%%%%%%%%%%%%%
\section{Kinematics of the broad line region}\label{sec5} 
\begin{figure}
\centering
\includegraphics[width=0.49\textwidth]{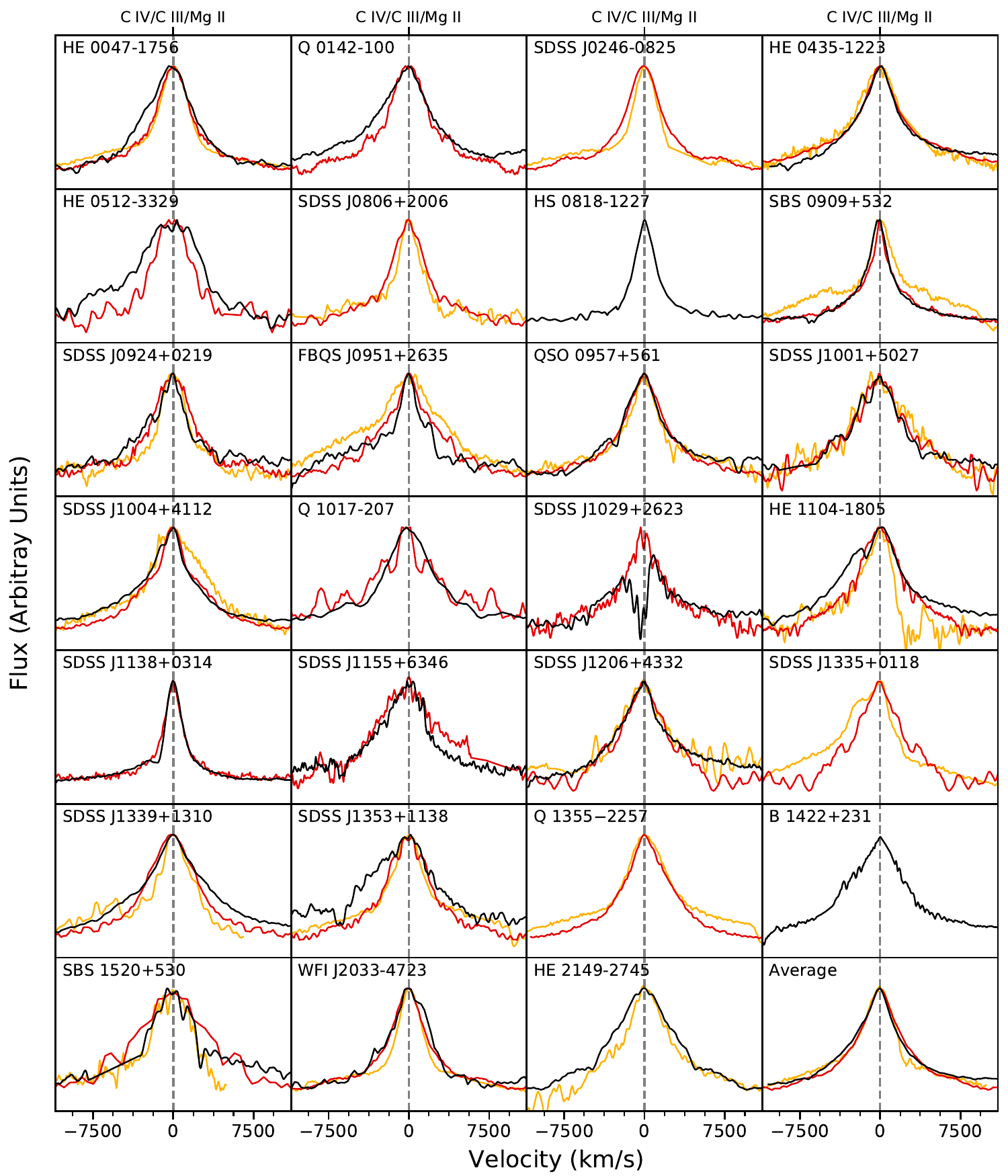}
\caption{Line profiles of C IV (black), C III] (red), and Mg II (yellow) as a function of velocity.}
\label{line_velocity_matched}
\end{figure}

We infer some results on the kinematics of the BLR by studying the line profiles corresponding to C IV, C III], and Mg II as a function of velocity.  In Figure \ref{line_velocity_matched} a superposition of the average line profiles for each system can be seen as well as the global average for all systems. The wings of Mg II correspond to relatively low velocities as compared with C III] and C IV, which is consistent with the weak impact of microlensing on this line. From Figure \ref{line_velocity_matched} we see that, in general\footnote{HE 0047-1756, SDSS J0246-0825, SDSS J0806+2006, SDSS J0924+0219, QSO 0957+561, HE 1104-1805, SDSS J1339+1310, SDSS J1353+1138, WFI J2033-4723, HE 2149-2745}, the core ($|\Delta v| \leq 700$ km/s) and the emission line component defined by $|\Delta v| < 3000$ km/s is narrower for Mg II when compared with those of C III] and C IV. The high-velocity wings of Mg II ($|\Delta v| > 4000$ km/s) are blended by several Fe II lines (\citealt{Vestergaard2001,Vestergaard2011}) and no reasonable comparison can be made. This is biasing the comparison of the velocity cores in several cases (e.g., SBS 0909+532 or FBQS J0951+2635).\\

Regarding C III], to ease visualization in the case of symmetric line profiles, we mirror the red part of the emission line profile in order to remove the blending by Al III and Si III in the blue wing. In most cases\footnote{HE 0047-1756, Q 0142-100, HE 0435-1223, SBS 0909+532, QSO 0957+561, SDSS J1001+5027, SDSS J1004+4112, HE 1104-1805, SDSS J1138+0314, SDSS J1206+5332}, the red part of C III] and C IV match very well, except at the lowest intensity level where the shelf-like feature $\sim\lambda1610$ (see, e.g., \citet{Fian2018}) is present in C IV. This striking kinematic coincidence propounds that both lines are mostly generated in the same region. However, the C IV line can be strongly affected by microlensing (see, for instance, the cases of QSO 0957+561, SDSS J1004+4112, SDSS J1206+4332 and SDSS J1339+1310), whereas C III] seems to be rather insensitive to this effect (except the blue wing of SDSS J1004+4112). This fact reveals (in agreement with previous studies; see, e.g., \citealt{Fian2018}) the existence of emitters located in a region small enough to be prone to microlensing, exclusively contributing to the C IV line but not to C III].\\

We made four rough estimates of the distance, $d$, moved by the accretion disk relative to the magnification pattern during the time elapsed between epochs of observation, $t_{obs}$ ($\sim$1 year in SDSS J1339+1310, $\sim$12 years in SDSS J1206+4332, $\sim$13 years in SDSS J1004+4112, and $\sim$17 years in QSO 0957+561), $d=v_{eff} \cdot t_{obs}$. The effective velocity for each source, $v_{eff}$, was estimated by dividing its Einstein Radius, $R_E$, by the Einstein crossing time, $t_E$, for this system. These values were taken from \citet{Mosquera2011}. As discussed in \citet{Fian2018}, the distance traveled in the source plane is too small ($\sim$0.6 light-days) for SDSS J1339+1310 to see variability in both wings. However, for SDSS J1206+4332 and SDSS J1004+4112 we obtained more interesting results, owing to larger displacements of the source ($\sim$8 light-days for SDSS J1206+4332 and $\sim$12 light-days for SDSS J1004+4112). If we compare the four epochs in SDSS J1004+4112, we can see that the strongly magnified blue wing fades while the red wing enhances. In SDSS J1206+4332 we detect, when comparing epoch I and II, a slight demagnification of image B in the blue wing while the red wing  enlarges. These observations confirm that the separation between the approaching and recending parts of the microlensed region of the BLR is about a few light-days in size. The case of QS0 0957 is more complex as the changes induced in the red wing may be due to intrinsic variability combined with the large time-delay between the images plus a possible contribution from microlensing. Although the distance moved on the source plane by this system is rather large ($\sim$17 light-days), we do not detect an enhancement in the blue wing. This result can be explained by an unusually large accretion disk (paper in preparation).\\

\begin{figure}
\centering
\includegraphics[width=0.45\textwidth]{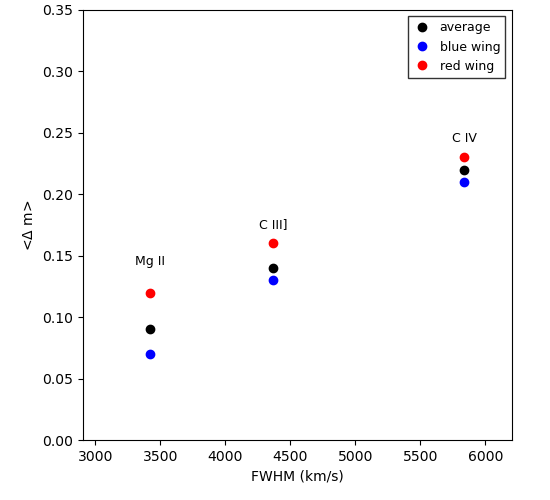}
\caption{Average amplitude of microlensing between images as a function of the line broadening for C IV, C III], and Mg II.}
\label{m_vs_fwhm}
\end{figure}

\begin{figure}
\centering
\includegraphics[width=0.47\textwidth]{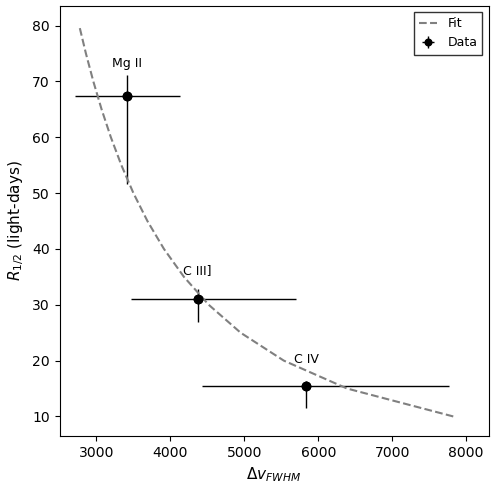}
\caption{Average emission region size as a function of the line broadening for C IV, C III], and Mg II. The line of best fit according to Eq. \ref{virial} is shown as a dashed gray line.}
\label{SMBH}
\end{figure}

In Figure \ref{m_vs_fwhm} we show the average amplitude of microlensing between images of all systems as a function of the Mg II, C III], and C IV line broadenings. By using the average line profile of as many 
objects as possible we hope to even out the deviations in the line 
profile introduced by individual objects. We do not take into account noisy spectra and spectra with absorption features. There are a couple of important issues to keep in mind when estimating the FWHM of the BELs: (i) it plays an important role where to fit and subsequently subtract the continuum underneath the line; for all systems and images we fit the continuum in the same wavelength regions (we allow some margin for a better adjustment of the fit to the corresponding continuum); (ii) blending by other emission lines may lead to an overestimation of the FWHM; (iii) an added complication in measuring the FWHM is the setting of the continuum level in the emission line. To estimate the uncertainties in the measured FWHM, we vary the zero point of the line by $\pm$ 10$\%$ (see Figure \ref{SMBH}), leading to variations of $\sim$25$\%$ in the FWHM. We note that one of the main sources of error in the measurement of the FWHM of C IV is the fraction of the shelf-like feature at $\sim\lambda1610$ under the emission line. The C III] is blended with Al III and Si III in the blue wing and therefore this wing cannot be used to compute the FWHM. It is important to remove these effects when measuring the line width to prevent an overestimation of the FWHM. We attempt this by using twice the distance of the half maximum of the line center to the red wing. In the case of Mg II, the line is located in the middle of the small blue bump of thousands of Fe II lines and is seriously contaminated by the blended iron lines and by the Balmer continuum (\citealt{Vestergaard2001,Vestergaard2011,Kova2017}). Hence, we are likely overestimating the FWHM of the Mg II line since half of the Mg II profile is submerged in the Fe II emission.

% %%%%%%%%%%%%%%%%%%%%%%%%%%%%%%%%%%%%%%%%%%%%%%%%%%%%%%%%%%%%%%%%%%%%%%%%%%%%%%%%%%%%%%%%%%
% %%%%%%%%%%%%%%%%%%%%%%%%%%%%%%%%%%%%%%%%%%%%%%%%%%%%%%%%%%%%%%%%%%%%%%%%%%%%%%%%%%%%%%%%%%
\section{Average estimate of the SMBH mass}\label{sec6}
It has become common practice to estimate central black hole masses based on the information that can be obtained from single-epoch spectra of multiple lensed quasars in a similar way as with reverberation mapping (RM). The Doppler broadening of the emission lines (FWHM) can be used as a proxy for the virial velocity (\citealt{Shen2012,Marziani2012,Coatman2016}) and combined with the size estimates of the emitting regions provided by microlensing we can estimate the virial mass, $M_{BH}$, according to
\begin{equation}
M_{BH}\simeq 9.8 \times 10^7 M_\odot\ f \left(\frac{R_{BLR}}{5{\rm\, light\, days}}\right) \left( {\frac{\Delta v_{FWHM}}{10000 {\rm \, km\, s^{-1}}}}\right)^2,
\label{virial}
\end{equation}\\
with $f$, the fudge factor, containing the unknown characteristics of structure, geometry, and details of the velocity field and its inclination along the line-of-sight (see \citealt{Peterson2011}). This approach assumes that the BLR is virialized and provides a powerful tool for obtaining black hole masses of distant quasars by using the emission line widths of Mg II and C IV to probe the BLR velocities when the Balmer lines are redshifted out of the optical observing window (\citealt{Vestergaard2001}).\\

We applied this relationship to the C IV, C III], and Mg II emission lines with $\Delta v_{FWHM}$ corresponding to the FWHM of the average line profiles obtained from our sample and $R_{BLR}$ corresponding to the sizes estimated using microlensing. We used a $\chi^2$ criterion to test for goodness of fit to those three points, and we obtain a black hole mass of $M_{BH} \sim 2.4_{-0.4}^{+1.5} \times 10^8 M_\odot$ for $f=2$, which is a reasonable result for the bright quasars of our sample (\citealt{Mosquera2011}). From Figure \ref{SMBH} we can conclude that, on average, the microlensing-based sizes are in agreement with the hypothesis of virialized kinematics. As discussed in Section \ref{sec4}, the size estimates of C III] and C IV could be biased toward small values. Hence, this mass estimate is, likely, an underestimate. Our result is consistent within the uncertainties with the black hole mass obtained in Paper I ($M_{BH} \sim 3.9_{-1.4}^{+1.8} 10^8 M_\odot$) by considering the Fe III $\lambda\lambda$2039-2113 blend of size $R_{BLR} \sim 11.3_{-4}^{+5}$ light-days and of velocity $\Delta v_{FWHM}$ $\sim$9400 km/s.

% %%%%%%%%%%%%%%%%%%%%%%%%%%%%%%%%%%%%%%%%%%%%%%%%%%%%%%%%%%%%%%%%%%%%%%%%%%%%%%%%%%%%%%%%%%
% %%%%%%%%%%%%%%%%%%%%%%%%%%%%%%%%%%%%%%%%%%%%%%%%%%%%%%%%%%%%%%%%%%%%%%%%%%%%%%%%%%%%%%%%%%
\section{Conclusions}\label{sec7}
In \citet{Fian2018} we studied microlensing in the wings of the BELs among 11 strongly lensed quasars; in this work we extend the study by 16 systems and analyze their spectra in up to five epochs. In total, taking into account all images and all epochs, we studied $\sim$100 C IV lines, more than 110 C III] lines, and around 80 Mg II lines. We have identified various signatures of microlensing in the wings of the BELs, and, subsequently, measuring its strength allowed us to constrain the sizes of their emitting region and hence study their kinematics. We finally discuss the consequences of our results and draw the following conclusions:\\

\begin{itemize}
\item[1 --]\textit{Microlensing.}
The main results of the previous sections are that the Mg II line is formed (as usually assumed) in an outer part of the BLR as it is only weakly affected by microlensing, whereas the wings of C IV suffer strong microlensing, indicating that this line forms very close to the accretion disk. We find in our sample that around 50$\%$ of the systems show microlensing in the C IV line. C III] is also significantly microlensed in about 25$\%$ of the systems. Significant variations caused by microlensing are rare in the wings of Mg II, with only two systems showing changes associated with microlensing. We do not detect microlensing simultaneously in the three emission lines, indicating a different origin/size of their emitting region.\\

\item[2 --]\textit{Intrinsic variability.}
The temporal sampling (up to five epochs of observation) allowed us to identify intrinsic variability and to classify the differences between pairs of spectra as candidates for intrinsic variability or microlensing. We derive robust estimates of both, intrinsic variability and amplitude of microlensing between the lensed images, and consistently separated a group of five objects dominated by microlensing (SBS 0909+532, FBQS J0951+2635, SDSS J1004+4112, SDSS J1206+4332, and SDSS J1339+1310) and another group of three systems in which intrinsic variability prevails (Q 0142-100, HE 1104-1805, and WFI 2033-4723). Two systems (HE 0435-1223 and HE 2149-2745) seem to be hybrid cases with both microlensing and intrinsic variability present. Finally, the case of QSO 0957+561 may be explained by intrinsic variability combined with the large time-delay between the images plus a possible contribution from microlensing.\\

\item[3 --]\textit{Size, structure, and geometry.} 
The fact that the wings of Mg II are generally not affected by microlensing whereas the wings of C IV (C III]) show strong (moderate) changes favors the hypothesis of two distinct regions, one large and insensitive to microlensing and the other one small and prone to microlensing. We frequently detect microlensing in either the blue or the red wing of C III] and/or C IV instead of a signal affecting symmetrically both wings. This result implies that the BLR does not have a spherically symmetric geometry.\\

From a statistical analysis using measured microlensing magnifications between image pairs of 27 lensed quasars we estimate the average sizes for the BELs from the product of individual likelihood functions for each image pair, epoch and system. This method includes the cases with little or no microlensing that by themselves contribute to the size estimates. Consistent with other recent studies (see \citealt{Guerras2013} and references therein) we found that microlensing depends on the degree of ionization, with magnifications more pronounced in the high-ionization lines. Taking as reference the cores (which have been considered unchanging in single epoch based studies), we find that the wings of Mg II are not significantly affected by either microlensing or intrinsic variability. The relative impact of microlensing is larger in C III] when compared with Mg II, indicating that the emission region associated with this line is more compact. We obtain size estimates of $\sim$31 and $\sim$67 light-days for the C III] and Mg II emission lines, in good agreement with RM studies (see, e.g., \citealt{Clavel1991,Homayouni2020,Wang2020}). The high impact of microlensing in the C IV emission line indicates that this line arises from a region of $\sim$16 light-days in size.\\

\item[4 --]\textit{Kinematics.}
From the impact of microlensing in the wings of C IV, C III], and Mg II, we attempt to broadly outline a basic relationship between kinematics and structure in the BLR. The unblended red part of the average line profiles of C IV and C III] match very well in the absence of microlensing. This kinematic coincidence indicates that both lines arise mainly from the same region. However, this alikeness in the line profiles is broken by the changes induced my microlensing, resulting in strong deformations in the C IV line profile while leaving unchanged the line profile of C III]. The high impact of microlensing in C IV reveals the existence of a second region (a few light-days in size), only contributing to the C IV line but not to C III] or Mg II. \\

\item[5 --]\textit{Line profile deformations.}
Depending on the structure of the BLR, microlensing could modify the broad line profiles. The frequent observation of asymmetric deformations of the emission lines and detection of microlensing in only one of the wings (i.e., either of the red or blue component) demonstrates that the BLR does not have, in general, a spherically symmetric geometry. Microlensing of a spherically symmetric BLR would lead to symmetric variations in the emission lines, while microlensing of a Keplerian disk leads to asymmetric variations (see, e.g., \citealt{Sluse2012} and references therein). We have been able to unveil outstanding microlensing-induced deformations in the wings of CIV in four systems (QSO 0957+561, SDSS J1004+4112, SDSS J1206+4332, and SDSS J1339+1310). For SDSS J1206+4332 and SDSS J1339+1310 we detect clearly asymmetrical enhancements toward the red. In the well studied case of SDSS J1004+4112 (\citealt{Richards2004,Gomez2006,Motta2012,Fian2016} and references therein) we detect variable enhancements in both wings, again highly asymmetric but this time the blue part is dominant. Intrinsic variability seems to affect the wings with similar strength, although no outstanding evidence of asymmetry associated with intrinsic variability has been detected. These simple observations support the hypothesis that the small region prone to both microlensing and intrinsic variability is intrinsically symmetrical, and that the asymmetry induced by microlensing in the line profile is related to the anisotropic spatial distribution of microlensing magnification at the source plane.\\ 

\item[6 --]\textit{SMBH mass.}
The microlensing-based sizes for the emitting regions of the BELs C IV, C III], and Mg II combined with the kinematic information (i.e., Doppler broadening of the emission lines) can be used to estimate the mass of the central SMBH. We obtain an average SMBH mass of $M_{BH}\sim 2.4_{-0.4}^{+1.5} \times 10^8 M_\odot$ for f = 2, which is consistent within the uncertainties with the black hole mass obtained in \citet{Fian2018} and thus, is a reasonable result for the bright quasars of our sample.
\end{itemize}

% %%%%%%%%%%%%%%%%%%%%%%%%%%%%%%%%%%%%%%%%%%%%%%%%%%%%%%%%%%%%%%%%%%%%%%%%%%%%%%%%%%%%%%%%%%
% %%%%%%%%%%%%%%%%%%%%%%%%%%%%%%%%%%%%%%%%%%%%%%%%%%%%%%%%%%%%%%%%%%%%%%%%%%%%%%%%%%%%%%%%%%

\begin{acknowledgements}
We thank the anonymous referee for the helpful comments and the constructive remarks on this manuscript. C.F. acknowledges the financial support the Tel Aviv University and University of Haifa through a DFG grant HA3555-14/1. E.M. and J.A.M are supported by the Spanish MINECO with the grants AYA2016- 79104-C3-1-P and AYA2016-79104-C3-3-P. J.A.M. is also supported from the Generalitat Valenciana project of excellence Prometeo/2020/085. J.J.V. is supported by the project AYA2017-84897-P financed by the Spanish Ministerio de Econom\'\i a y Competividad and by the Fondo Europeo de Desarrollo Regional (FEDER), and by project FQM-108 financed by Junta de Andaluc\'\i a. V.M. acknowledges partial support from Centro de Astrof\'{\i}sica de Valpara\'{\i}so. Research by D.C. has been partially supported by an Israeli Science Foundation grant 2398/19.\\
\end{acknowledgements}

\bibliographystyle{aa}
\bibliography{bib_paper}

\end{document}